\documentclass[%
reprint,
superscriptaddress,
notitlepage,
amsmath,amssymb,
prf,
onecolumn,
floatfix,
tightenlines,
longbibliography,
11pt,
]{revtex4-1}
\usepackage{natbib}
\usepackage{subfigure}

\usepackage[normalem]{ulem}
\usepackage{graphicx}
\usepackage{xcolor}
\usepackage{epstopdf}
\usepackage{appendix}
\usepackage{amsmath}
\usepackage{amsfonts}
\usepackage{comment}

\mathchardef\arr="017E % character 7E from textfont 1 is the vector arrow 

\def\w {\mathbf{w}}
\def\x {\mathbf{x}}
\def\n {\hat{\mathbf{n}}}
\def\p {\hat{\mathbf{p}}}
\def\e {\hat{\mathbf{e}}}
\def\f {\mathbf{f}}
\def\T {\mathbf{T}}
\def\g {\mathbf{g}}

\begin{document}

\title{Experimental validation of fluid inertia models for a cylinder settling in a quiescent flow}
\author{F. Cabrera}
\affiliation{Univ Lyon, ENS de Lyon, Univ Claude Bernard Lyon 1, CNRS, Laboratoire de Physique, F-69342 Lyon, France}
\author{M.Z. Sheikh}
\affiliation{Univ Lyon, ENS de Lyon, Univ Claude Bernard Lyon 1, CNRS, Laboratoire de Physique, F-69342 Lyon, France}
        \affiliation{Department of Mechanical Engineering, University of Engineering and Technology, 54890, Lahore, Pakistan}
\author{B. Mehlig}
\affiliation{Department of Physics, Gothenburg University, 41296 Gothenburg, Sweden}
\author{N. Plihon}
\affiliation{Univ Lyon, ENS de Lyon, Univ Claude Bernard Lyon 1, CNRS, Laboratoire de Physique, F-69342 Lyon, France}
\author{M. Bourgoin}
\affiliation{Univ Lyon, ENS de Lyon, Univ Claude Bernard Lyon 1, CNRS, Laboratoire de Physique, F-69342 Lyon, France}
\author{A. Pumir}
\affiliation{Univ Lyon, ENS de Lyon, Univ Claude Bernard Lyon 1, CNRS, Laboratoire de Physique, F-69342 Lyon, France}
\affiliation{Max-Planck Institute for Dynamics and Self-Organization, G\"ottingen, D-37077, Germany}
\author{A. Naso}
\affiliation{Univ Lyon, CNRS, Ecole Centrale de Lyon, INSA Lyon, Univ Claude Bernard Lyon 1, Laboratoire de M\'ecanique des Fluides et d'Acoustique, UMR 5509, 69130 Ecully, France}

\begin{abstract}
The precise description of the motion of anisotropic particles in a flow rests on the understanding of the force and torque acting on them. Here, we study experimentally small, very elongated particles settling in a fluid at small Reynolds number. In our experiments, we can, to a very good approximation, relate the rate of rotation of cylindrical tungsten rods, of aspect ratios $\beta = 8$ and $\beta = 16$, settling in pure glycerol to the torque they are experiencing. This allows us to compare the measured torque with expressions obtained either in the slender-rod limit, or in the case of spheroids. Both theories predict a simple angle dependence for the torque, which is found to capture very well the experimental results. Surprisingly, the slender-rod approximation predicts much better the results for $\beta = 8$, than for $\beta = 16$. In the latter case, the expression obtained for a spheroid provides a better approximation. The translational dynamics is shown to be in qualitative agreement with the slender-rod and spheroid models, the former one being found to represent better the experimental data.
\end{abstract}

\maketitle

\section{Introduction}
The settling of non-spherical particles at low Reynolds number in a fluid at rest is a subtle problem. It has been known for a long time that describing the angular degrees of freedom with the Stokes approximation leads to an indeterminacy in the settling angle: the particle may settle at any angle with respect to the vertical. This degeneracy, however, is lifted by the action of the fluid, already in the limit
of a small but non-zero particle Reynolds number ${\rm Re}_p$. Whereas the correction to the translational velocity has been understood for a long time,  the torque acting on an non-spherical particle is not as well understood. Cox~\cite{cox1965steady} determined the torque acting on particles. Later, Khayat and Cox~\cite{khayat1989inertia} determined the torque for slender particles. More recently, the torque acting on spheroids was determined for spheroids of arbitrary dimensions ~\cite{dabade2015effects}. These explicit analytic expressions are very useful to describe the motion of particles at small Reynolds numbers.

Because of the importance of the problem, many contributions have been devoted to its numerical solution, to determine approximate torque parameterizations as a function of the particle Reynolds number ${\rm Re}_p$ and particle shape~\cite{Hoelzer09,Zastawny12,Jiang14,Ouchene15,Ouchene16,Andersson19,Froehlich20}. 
The effect of small to moderate Reynolds numbers for spheroids over a large range of aspect ratios $\beta$ was recently analysed by means of numerical simulations~\cite{jiang2020inertial}, allowing a direct comparison with the small-${\rm Re}_p$ theory~\cite{dabade2015effects} in the limit of small ${\rm Re}_p$.
It was found that the theory~\cite{dabade2015effects} generally provides the correct functional form and the correct magnitude for the torque. However, it tends to over predict the numerically determined values when ${\rm Re}_p \gtrsim 1$.
The hydrodynamic loads on a fixed finite-length circular cylinder in a uniform flow have also been obtained numerically, from creeping-flow conditions to strongly inertial regimes \cite{Kharrouba21}. Semi-empirical models based on theoretical predictions and incorporating finite-length and inertial corrections extracted from the numerical data were then derived. There have been, in comparison, far fewer experimental studies. 

In experiments where rods settle through a simple vortical flow, it was shown that the effect of fluid inertia could not be ignored~\cite{lopez2017inertial}. 
 More recently, the force and torque measured in settling experiments of symmetric and asymmetric fibers were accurately predicted by the model of Khayat and Cox~\cite{roy2019inertial}. The expression of torque by Khayat and Cox for symmetric fibers overpredicts the torque experimentally measured for $\beta=20$ and ${\rm Re}_p=1.6$, while it underpredicts the measurements for $\beta=100$ and ${\rm Re}_p=8.6$ -- however the origin was argued to lie in the defects in fibers which produce slight asymmetries~\cite{roy2019inertial,candelier_mehlig_2016}.\\
Here, we consider again the problem in a different setup, using small tungsten rods, roughly $ 12 $ times denser than the surrounding fluid, which is taken as pure glycerol. The resulting particle Reynolds number, $U_s l/\nu$, where $l$ is the half-length of the rod, $U_s$ the settling velocity and $\nu$ the fluid viscosity, is always smaller than 0.39. Our experimental setup (based on 3D tracking of the particle rotation and translation) allows us to determine precisely all the degrees of freedom of the motion, and to deduce from it the torque acting on the particle.
Our experimental results are then systematically compared to the predictions for slender-rods and spheroids.
 
One motivation for studying the problem comes from the modeling of processes involving non-spherical particles settling in turbulent flows. This includes, in an engineering context, problems involving paper fibers~\cite{lundell2011fluid}. The question is also particularly relevant in the environmental sciences, such as the settling of planktons in the oceans~\cite{durham2013turbulence,pedley1992hydrodynamic,ruiz2004turbulence}, or ice crystals settling in clouds~\cite{pruppacher1980microphysics,siewert2014orientation,gustavsson2017statistical,jucha2018settling,
gustavsson2019effect,sheikh2020}.

The article is organized as follows. We first introduce in Sec. \ref{sec:eq} the translational and rotational equations of motion for spheroids and slender-rods. The experimental design is then described in Sec. \ref{sec:exp_design}. The results are presented in Sec. \ref{sec:results}. Finally, our conclusions are summarized in Section \ref{sec:conclusions}.
%%%%%%%%%%%%%%%%%%%%%%%%%%%%%%%%%%%%%%%%%%%%%%%%%%%%%%
%%%%%%%%%%%%%%%%%%%%%%%%%%%%%%%%%%%%%%%%%%%%%%%%%%%%%%
\section{Theoretical background} \label{sec:eq}
\begin{figure}[t]
\centering
%\begin{overpic}
\includegraphics[width = 0.4\textwidth]{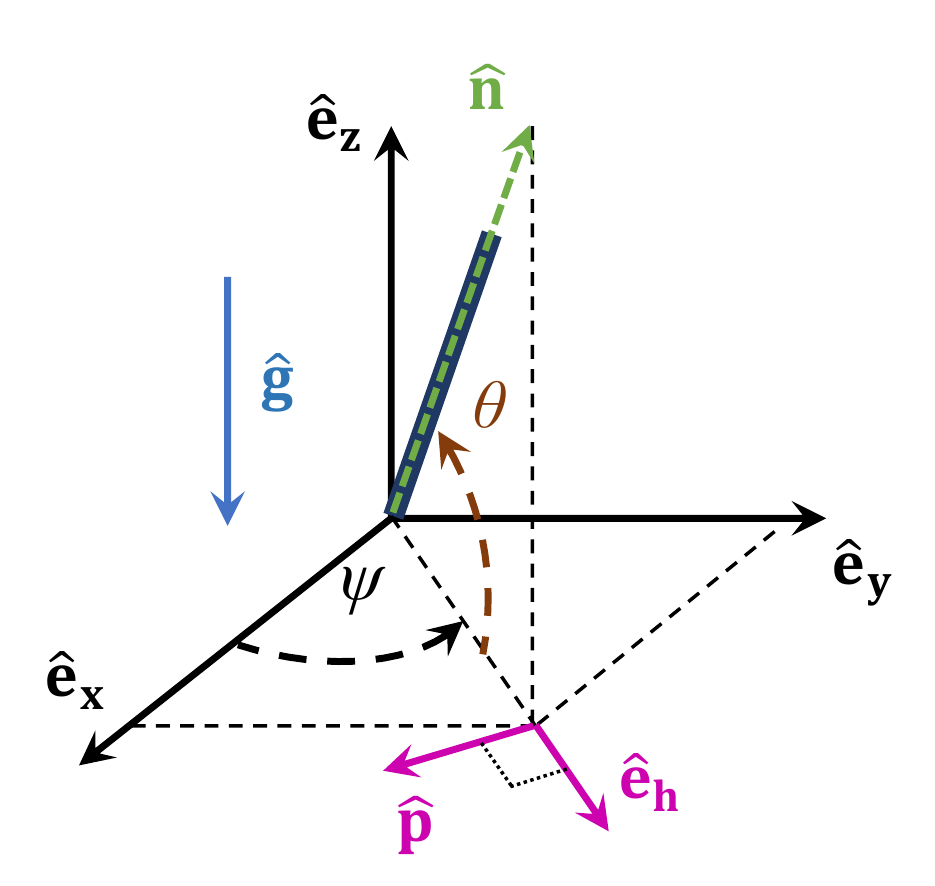}
\hspace{1cm}
\includegraphics[width = 0.4\textwidth]{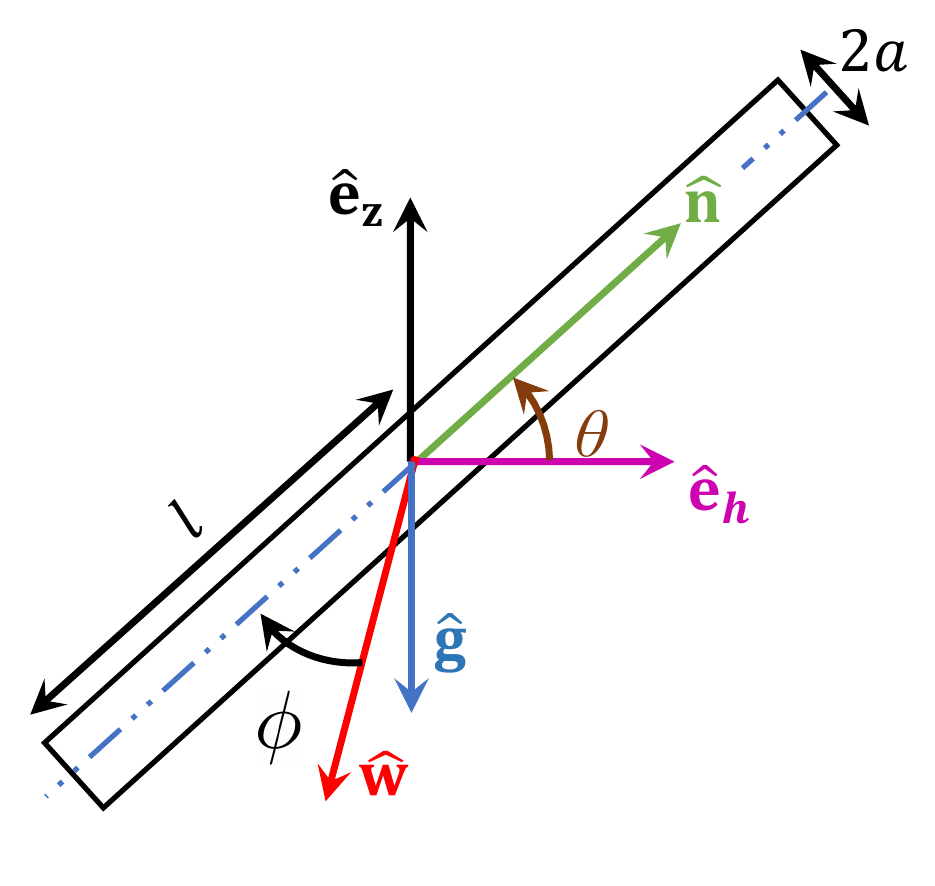}
%trim={<left> <lower> <right> <upper>}
\put (-400,160) {\large{\textbf{(a)}}}
\put (-200,160) {\large{\textbf{(b)}}}
%\end{overpic}
\caption{Representation of the settling cylindrical particle (a) in 3D and (b) in the vertical plane $(\n,\g)$. See text for details.}
\label{fig:notation}
\end{figure}

%%%%%%%%%%%%%%%%%%%%%%%%%%%%%%%%%%%%%%%%%%%%%%%%%%%%%%
%%%%%%%%%%%%%%%%%%%%%%%%%%%%%%%%%%%%%%%%%%%%%%%%%%%%%%
Our experiments were performed with cylindrical rods of density $\rho_p$, half-length $l$ and radius $a$. The particle aspect ratio is $\beta = l/a$.
The motion of the particles is characterized by the center of mass, $\x$, and 
the unit vector, $\n$, characterizing the orientation, as illustrated in 
Fig.~\ref{fig:notation}. The orientation of $\n$
is defined by the angles $\psi$ and $\theta$.
The vector $\p$ is orthogonal to the projection of $\n$ onto the horizontal $(\e_x,\e_y)$
plane, $\p = \sin(\psi) \e_x - \cos(\psi) \e_y$, and $\e_z$ is the vertical. 
In the experiments, the initial angular velocity was orthogonal to the plane
spanned by $\g$ and $\n$ (See Section \ref{sec:results}). In other words, 
the angle $\psi$ was found to remain constant, so the angular dynamics 
can be described by $\theta$ only.

The particle Reynolds number is defined as
\begin{equation}
{\rm Re}_p = \frac{ l |\w|}{\nu},
\label{eq:Re}
\end{equation}
where $\w$ is the velocity of the center of mass. The particle Reynolds number allows us to quantify the respective role of the inertial and viscous effects
in the problem. In our experiments, ${\rm Re}_p$ does not exceed $1$. As a 
consequence, the force and the torque acting on the particle 
originate from the viscous term (the Stokes forces and torques, 
$\f_S$ and $\T_S$), plus contributions due to fluid inertia, which are
derived in a systematic perturbation expansion in the parameter ${\rm Re}_p$,
denoted here as $\f_I$ and $\T_I$.

The equations of motion for the translational degrees of freedom  read as
\begin{equation}
\frac{ d \x}{dt} = \w ~~~~ {\rm and } ~~~ \frac{d \w}{dt} = \g + \frac{\f_H}{m_p}, 
\label{eq:Newton_x}
\end{equation}
where $\f_H \equiv \f_S + \f_I$ is the total hydrodynamic force acting on the 
particle, and $m_p$ is its mass. Note that the effect of buoyancy can be simply taken into 
account by
the transformation: $\g \rightarrow \g (1 - \rho_f/\rho_p)$.
The general expression for the Stokes force is:
$\f_S = 6 \pi a \mu  \mathbb{A}(\n) (\mathbf{u} - \w)$, where $\mu$
is the dynamic viscosity of the fluid, $\mathbf{u}$ is
the undisturbed velocity of the fluid at the particle position (equal to zero in a quiescent flow),
and $\mathbb{A}(\n)$ is the resistance tensor~\cite{Kim:2005}. The tensor 
$\mathbb{A}$ can be expressed in terms of two coefficients, $A_{\perp}$ and 
$A_{\|}$: $A_{ij} = A_{\perp} ( \delta_{ij} - n_i n_j) + A_{\|} n_i n_j$.
The expressions of $A_{\perp}$ and $A_{\|}$ depend on the shape of the particle (rod or spheroid), and in particular on its aspect ratio $\beta$, and are given in Appendix~\ref{appendixA}. In the limit $\beta \gg 1$, the expressions reduce, for the two shapes considered, to $A_{\perp} = 2 A_{\|} = (4/3) [\beta/\log(\beta)]$. 
Equation~\eqref{eq:Newton_x}, together with the specific form of $\f_S$, 
imply that if $|\f_I|\ll |\f_S|$ the particle velocity $\w$ relaxes with a characteristic
(Stokes) time $\tau_p = a^2 \log(\beta) ( \rho_p/\rho_f)/(3 \nu)$. In our 
experiments, we observe that the characteristic time scales of the particles (of the order of seconds)
are much longer than $\tau_p$ (of the order of a milliseconds). This implies that the dynamics is overdamped, and that the center-of-mass velocity can be obtained by solving $\g + \f_H/m_p = 0$ at any time. 

The effect of finite fluid-inertia leads to a correction to the Stokes force.
In the case of a spheroid, this correction to the resistance tensor can 
be expressed as:
$\mathbb{A}^s  \rightarrow \mathbb{A}^s + \mathbb{A}^s_I$. 
Similarly, the expressions for the corrections to the force for a slender rod, $\mathbf{f}^r_I$, can 
be found in~\cite{khayat1989inertia}, and are given in the Appendix~\ref{appendixA}.\\
%%%%%%%%%%%%%%%%%%%%%%%%%%%%%%%%%%%%%%%%%%%%%%%%%%%%%%
%%%%%%%%%%%%%%%%%%%%%%%%%%%%%%%%%%%%%%%%%%%%%%%%%%%%%%
%%%%%%%%%%%%%%%%%%%%%%%%%%%%%%%%%%%%%%%%%%%%%%%%%%%%%%
%%%%%%%%%%%%%%%%%%%%%%%%%%%%%%%%%%%%%%%%%%%%%%%%%%%%%%
As stated before, in our experiments the change in the orientation of a particle is due to 
variations of the angle
$\theta$ only. The equation of motion for this angle is obtained by projecting
the equation for the angular momentum along the $\p$-direction:
\begin{equation}
I_p \ddot{\theta} = ( \T_S + \T_I) \cdot \hat{\mathbf{p}},
\label{eq:Newton_a}
\end{equation}
where $I_p$ is the moment of inertia of the particle with respect to its center
of mass, perpendicular to its axis ($I_p = \frac{1}{12} ml^2$). As it was the case for
the force, the torque can be written as the sum of a contribution due to 
viscous forces (Stokes), $\T_S$, and a contribution due to inertia, $\T_I$.
The general expression for the Stokes torque is~\cite{jeffery1922motion}:
$\T_S = 6 \pi a \mu [ \mathbb{C} (\boldsymbol{\Omega} - \boldsymbol{\omega} )  + \mathbb{H}: \mathbb{S} ]$,
where $\boldsymbol{\omega}$ is the angular velocity of the solid, $\boldsymbol{\Omega} = 1/2 (\nabla \wedge \mathbf{u})$ the vorticity, and $\mathbb{C}$ is the resistance tensor. 
The last term, $\mathbb{H}: \mathbb{S}$, involves the strain in the fluid, 
which is $0$ in a fluid at rest, as considered here.  In a quiescent fluid, 
the torque is therefore proportional to the particle angular velocity $\omega$ which reads $\dot{\theta}$ since the $\psi$ angle is constant:
\begin{equation}
T_S \equiv \T_S \cdot \p = - C_S \dot{\theta},
\label{eq:T_S}
\end{equation}
where $C_S$ is a resistance coefficient proportional to $\mu a^3$. The
dimensionless constant of proportionality depends on the details of the particle.
In the very large $\beta$ limit, the expression for $C_S$ reduces to
\begin{equation}
C_S = \frac{8 \pi}{3} \frac{\mu l^3}{\log \beta}.
\end{equation}
As it was the case for the dynamics of the center of mass, the resistance 
term $-C_S\dot{\theta}$ provides a characteristic relaxation time scale for the particle orientation which is, up to a numerical factor of order $1$, equal to the characteristic
time $\tau_p$ already introduced~\cite{Ana20}.
The time scale over which the angle $\theta$ evolves is also found to be
very long compared to $\tau_p$, which ensures that, to a very good 
approximation, the left-hand side of Eq.~\eqref{eq:Newton_a} can be set to $0$.
This allows us to relate directly, via Eq.~\eqref{eq:T_S}, the projection
of the torque due to the inertial forces, $\T_I$, perpendicularly to $\n$,
$T_I \equiv \T_I \cdot \p$:
\begin{equation}
T_I =  C_S \dot{\theta}.
\label{eq:TI_theta_dot}
\end{equation}
Determining the torque $\T_I$ is a very challenging problem, even in the small-${\rm Re}_p$ limit.

We compare our experimental results, carried out with rods,
to the results of the slender-rod theory of~\cite{khayat1989inertia} ,
valid asymptotically when $\beta^{-1} {\rm Re}_p \ll 1$ and $\beta \gg 1$.
The first criterion is very well satisfied in the experiments. The second one
is not, since $\beta = 8$ or $16$ in our experiments, and since the small
parameter in the slender-rod theory is $1/\log \beta$. We therefore compared
also with a second perturbative theory for the torque, valid when the particle
has a spheroidal shape~\cite{dabade2015effects}. This theory is valid for any arbitrary aspect
ratio, $\beta$, but only to leading order in ${\rm Re}_p$. Furthermore, it was found
numerically to describe the torque acting on spheroidal particles
quite well, with an accuracy of $\sim 20\%$ for $\beta = 6$ over the range
of ${\rm Re}_p$ in the experiment~\cite{jiang2020inertial}.
We emphasize that the theory was derived for spheroidal particles, but we 
expect that it nevertheless works qualitatively for rod-like particles of
the same aspect ratio and the same particle mass.

At small Reynolds numbers, symmetry 
considerations~\cite{subramanian2005inertial,jiang2020inertial} indicate that 
the torque takes the form
\begin{equation}
\T_I = F(\beta) \rho_f l^3 (\w \times \n) (\w \cdot \n).
\label{eq:T_symmetry}
\end{equation}
The expression \eqref{eq:T_symmetry} shows that $\T_I$ is along $\p$, and that its norm reads
\begin{equation}
T_I = F(\beta) \rho_f l^3|{\bf w}|^2 \sin \phi \cos \phi =  C_I | \w |^2 \sin 2\phi.
\label{eq:torque_inertial}
\end{equation}
The shape factor $F(\beta)$ depends only on the particle aspect ratio.
As a result, the factor $C_I$ is
known for spheroids~\cite{dabade2015effects} 
and in the slender-rod limit~\cite{khayat1989inertia,lopez2017inertial}; we denote its values as $C_I^s$ and $C_I^r$. The
corresponding formulae are summarized in Appendix~\ref{appendixB}. 
We note that the expressions for the two theories coincide in the limits ${\rm Re}_p \rightarrow 0$ and
$\beta \rightarrow \infty$, where: $C_I(\beta) \approx -  5 \pi/[ 3 (\log \beta)^2 ]\times (\rho_f l^3/2) $.
Combining Eqs.~\eqref{eq:TI_theta_dot} and \eqref{eq:torque_inertial}
leads to the prediction of the model that
\begin{equation}
\dot{\theta} = \frac{ C_I  }{C_S} | \w | ^2\sin 2 \phi.
\label{eq:prediction}
\end{equation}
This overdamped torque model for rods was found to 
qualitatively reproduce experimental results of rods settling in a cellular 
flow~\cite{lopez2017inertial}, and allowed to investigate the settling of anisotropic 
particles in turbulent flows~\cite{Kramel:PhD}. The expressions of the torque
for spheroids have been validated numerically~\cite{jiang2020inertial},
and used to study theoretically and numerically the settling
of spheroids in 
turbulence~\cite{gustavsson2019effect,sheikh2020,Ana20,Gus21}.
We notice that when the fluid is in motion, the local velocity gradients 
may contribute to the torque expression ~\cite{Einarsson2015}. 
In our problem, with $\rho_p/\rho_f \gg 1$, the corresponding contributions are negligible.

\section{Experimental setup}\label{sec:exp_design}
\begin{figure}
\centering
\includegraphics[scale = 0.3,trim={21cm 0 0 7.5cm},clip]{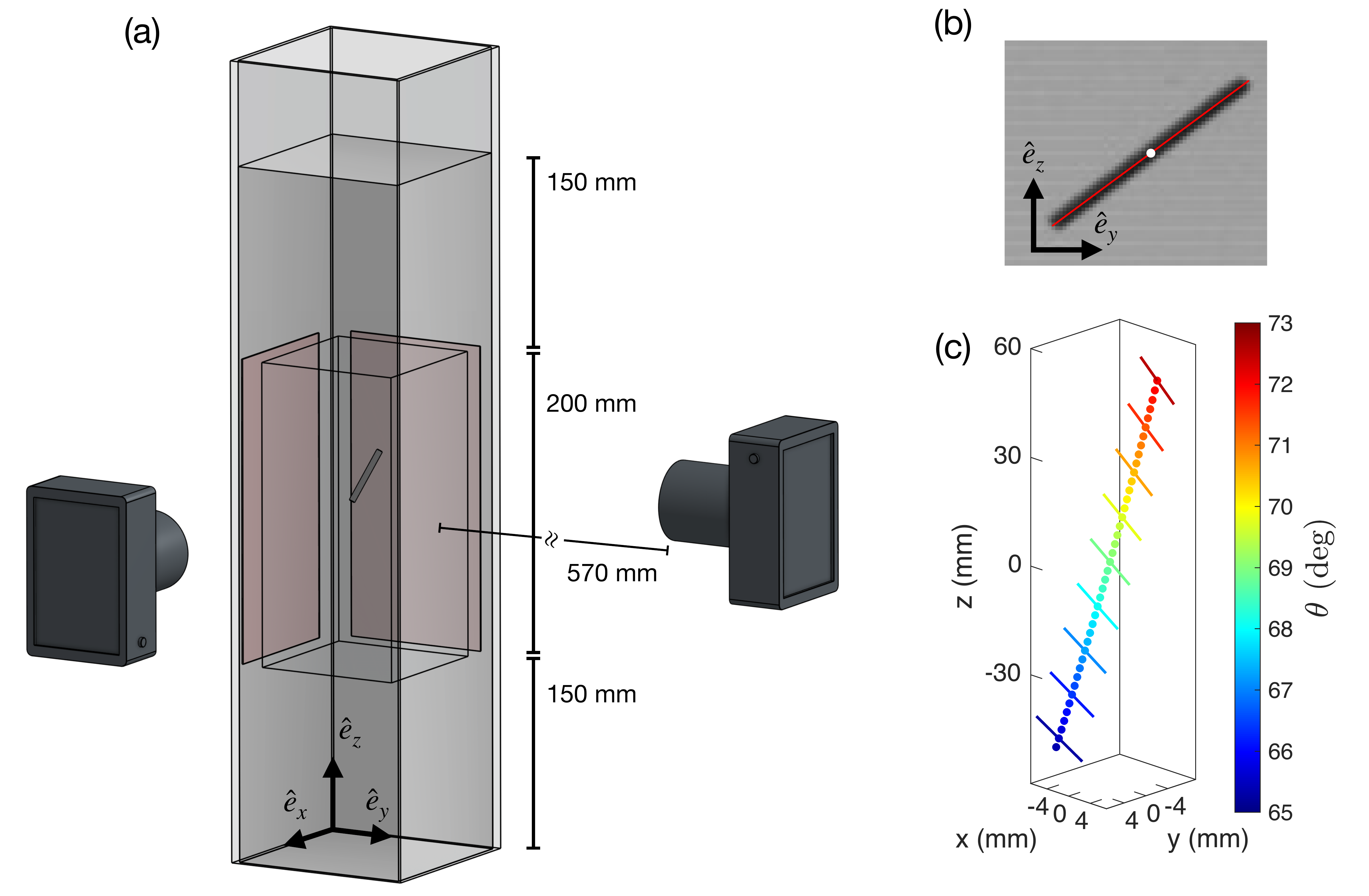}
\caption{(a) Schematic view of the experimental setup. The origin (0,0,0) is set at the geometric center of the vessel. (b) Typical raw image obtained from one camera showing the particle of aspect ratio $\beta =16$, the detected particle center line (in red) and center of mass (white dot). 
(c) Successive positions of a rod's center line and center of mass during one experiment (for $\beta =16$), color coded by the value of the angle $\theta$. For the sake of clarity, only one center of mass data-point out of 80, and one center line each 5 center of mass data-points are shown.
}
\label{fig:fiber}
\end{figure}
The experiments were performed in a PMMA tank with a square cross-section of side $150$~mm and height of $710$~mm (see Fig.~\ref{fig:fiber}~(a)). The tank was filled with pure glycerol (Sigma-Aldrich W252506-25KG-K), with a density $\rho_f = 1216$~kg~m$^{-3}$ and dynamic viscosity $\mu=1.05$~Pa.s at 22$^{\circ}$C in which we observed the settling of heavy cylindrical particles. The cylindrical particles were made of Tungsten-Carbide (WC with $5\%$ Cobalt), manufactured by Comac Europe, with a density $\rho_p = 14800$~kg/m$^{3}$, resulting in a fluid-to-particle density ratio of $\rho_p/\rho_f = 12.1$. Cylindrical rods were milled and filed at lengths $2l = (16.0\pm0.5)$~mm and ($8.0\pm0.5$)~mm, from long cylinders of diameter $2a=1$~mm, resulting in cylindrical particles with aspect ratios $\beta$ of 16 and 8. 
The surface roughness was measured using a Scanning Electron Microscope, leading to an average arithmetic roughness value of $15$~$\mu$m.   
The particle Reynolds number $Re_p$ was measured to lie between 0.255-0.390 for $\beta= 16$, and 0.105-0.150 for $\beta = 8$.
The particles trajectories were reconstructed from images acquired using two cameras (fps1000 model from The Slow Motion Company), placed orthogonal, as shown in Fig.~\ref{fig:fiber}(a). The cameras acquire images at a frame rate of $1400$~fps with a resolution of $720\times1280$~px$^2$. The volume over which the trajectories of the particles are reconstructed is $130\times130\times200$~mm$^3$ to avoid wall effects~\cite{walleffect} -- the distance between the rods and the walls is thus always larger than $10$~mm. 

The particles are backlight-illuminated by using two white and homogeneous light panels (light-red rectangles in Fig.~\ref{fig:fiber}~(a)). A typical image captured from one camera shows the projection of the cylindrical particle in the $(y,z)$ plane in Fig.~\ref{fig:fiber}(b). For each image, and on both cameras, the detection of the particle is computed using the MATLAB function \textit{regionprops}, which detects the particle’s major axis and minor axis, center of mass and orientation (see Fig.~\ref{fig:fiber}(b)). 
The 3D orientation of the particle is then obtained from a 3D Particle Velocity Tracking (PTV) algorithm~\cite{micaPTV}: a set of five points on the major axis are matched in 3D from the projections in the $(y,z)$ and $(x,z)$ planes (the five points are equally spaced between the two extrema). The particle position and orientation is thus reconstructed in 3D, as shown in Fig.~\ref{fig:fiber}(c), and all parameters related to the dynamics of the particles can be subsequently computed. The evolution of a typical sedimentation experiment is displayed  in  Fig.~\ref{fig:fiber}(c), from top to bottom as time increases, clearly demonstrating that the angle $\theta$ decreases in time.
The tracking error of the PTV system was shown to be inferior to $130~\mu$m, or equivalently $50\%$ of the pixel size. Additionally, the raw data was filtered via the convolution with a Gaussian kernel of variance $\sigma= 1.2\times10^{3}$ and $\sigma = 8\times10^{2}$, for $\beta = 8$ and 16 respectively. Note that to avoid perspective distortion of the projected contour of an anisotropic object~\cite{HUANG96, toupoint2019}, the cameras were installed sufficiently far from the sedimentation tank. 
An air-conditioning system keeps a constant room temperature at (22 $\pm$ 1)$^{\circ}$C, which bounds the viscosity variations  to 5$\%$. Moreover, to reach thermal equilibrium, a 48-hours delay was systematically respected between the filling of the tank with pure glycerol and the experiments. The calibration of the PTV system then consisted on the displacement of a target with known dimensions over the visualization volume~\cite{micaPTV}. The particles were released in the fluid with chemical tweezers: we completely submerged the particle and released it when the glycerol's free surface was at rest (approximately after $15$~s). A minimum time of $90$~s is taken between two successive realisations to assure that the fluid has no motion left from the previous drop. 
For both $\beta = 8$ and $\beta = 16$, 25 independent realisations have been acquired.   \\
With the values of the physical parameters in the experiment, the characteristic time scale of the particles, $\tau_p$, introduced in Section~\ref{sec:eq}, does not exceed $10^{-3}$ s, which, as we will document, is very short compared to the characteristic time of the evolution. Note that, additionally, there is $150$~mm of fluid above and below the visualization volume. A particle then travels for approximately 1000$\tau_p$ before and after it enters the detection volume: the fluid above makes the particle loose memory of any transient produced in the release; whereas the fluid below keeps the particle away from the bottom-wall at all times.

\section{Results}\label{sec:results}

Before we proceed to present our results, we recall that the motion of the rods was found to be, to a very good approximation, planar.
Namely, the variations of the angle $\psi$ between the projection of the ${\bf n}$ vector on a horizontal plane and an arbitrary fixed horizontal vector, as illustrated in Fig.~\ref{fig:notation}~(a), are less than $5^\circ$ during a given experiment, with a typical value of 2$^\circ$ over all realisations.
We therefore used the values of the torques valid in the case of a planar motion, where the orientation is parametrized solely by the angle $\theta$.
%%%%%%%%%%%%%%%%%%%%%%%%%%%%%%%%%%%%%%%%
%%%%%%%%%%%%%%%%%%%%%%%%%%%%%%%%%%%%%%%%
\subsection{Translational motion}
%%%%%%%%%%%%%%%%%%%%%%%%%%%%%%%%%%%%%%%%
%%%%%%%%%%%%%%%%%%%%%%%%%%%%%%%%%%%%%%%%
%%%%%%%%%%%%%%%%%%%%%%%%%%%%%%%%%%%%%%%%
%%%%%%%%%%%%%%%%%%%%%%%%%%%%%%%%%%%%%%%%
\begin{figure}
\centering
\begin{subfigure}{}
%\centering
%\begin{overpic}
\includegraphics[width=0.45\linewidth,trim={0 0 3.5cm 3cm},clip]{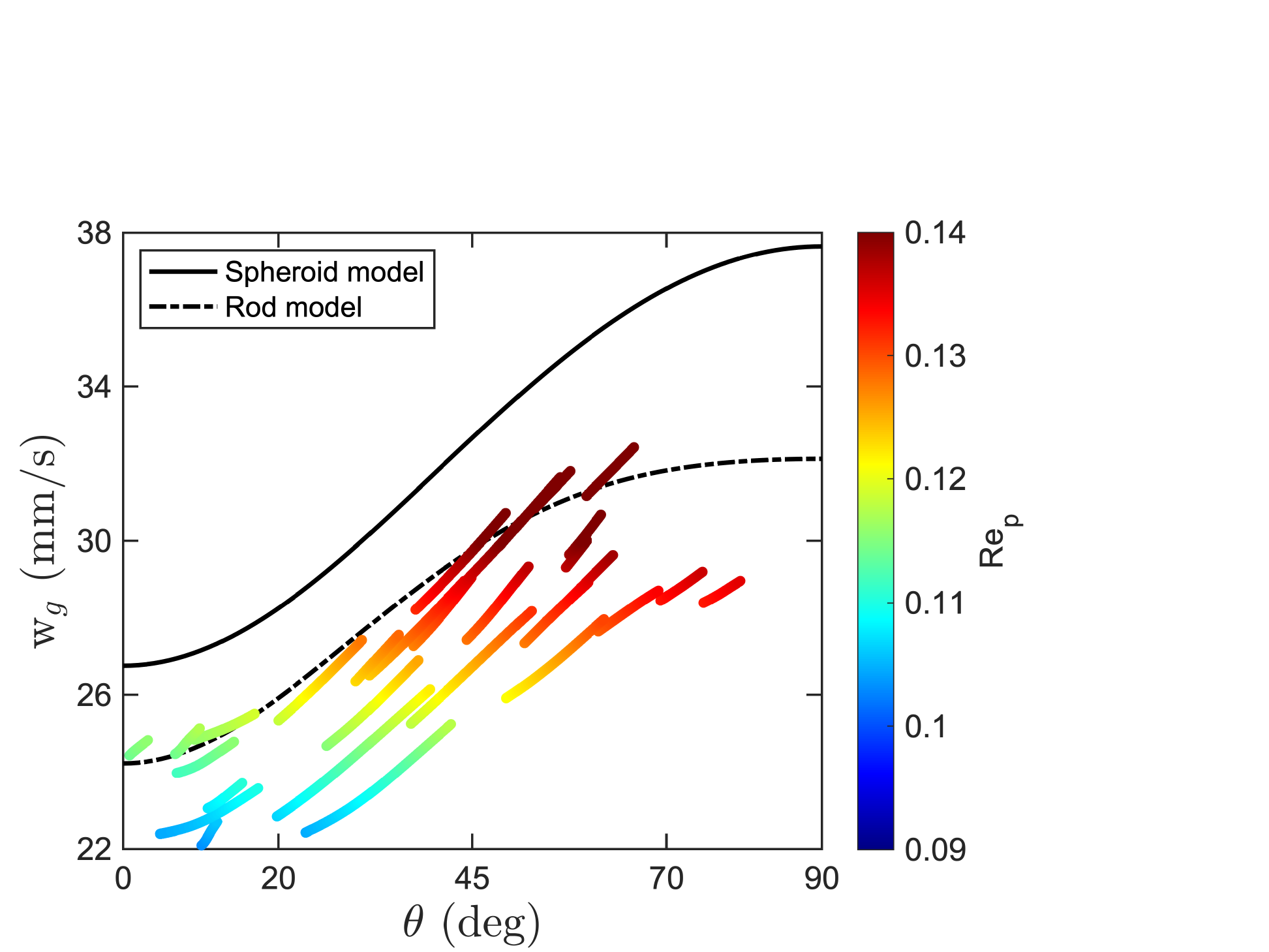}
\put (-130,155) {\large$\beta=8$}
\put (-223,150) {(a)}
%\end{overpic}
\end{subfigure}
\begin{subfigure}{}
%\centering
%\begin{overpic}
\includegraphics[width=0.45\linewidth,trim={0 0 3.5cm 3cm},clip]{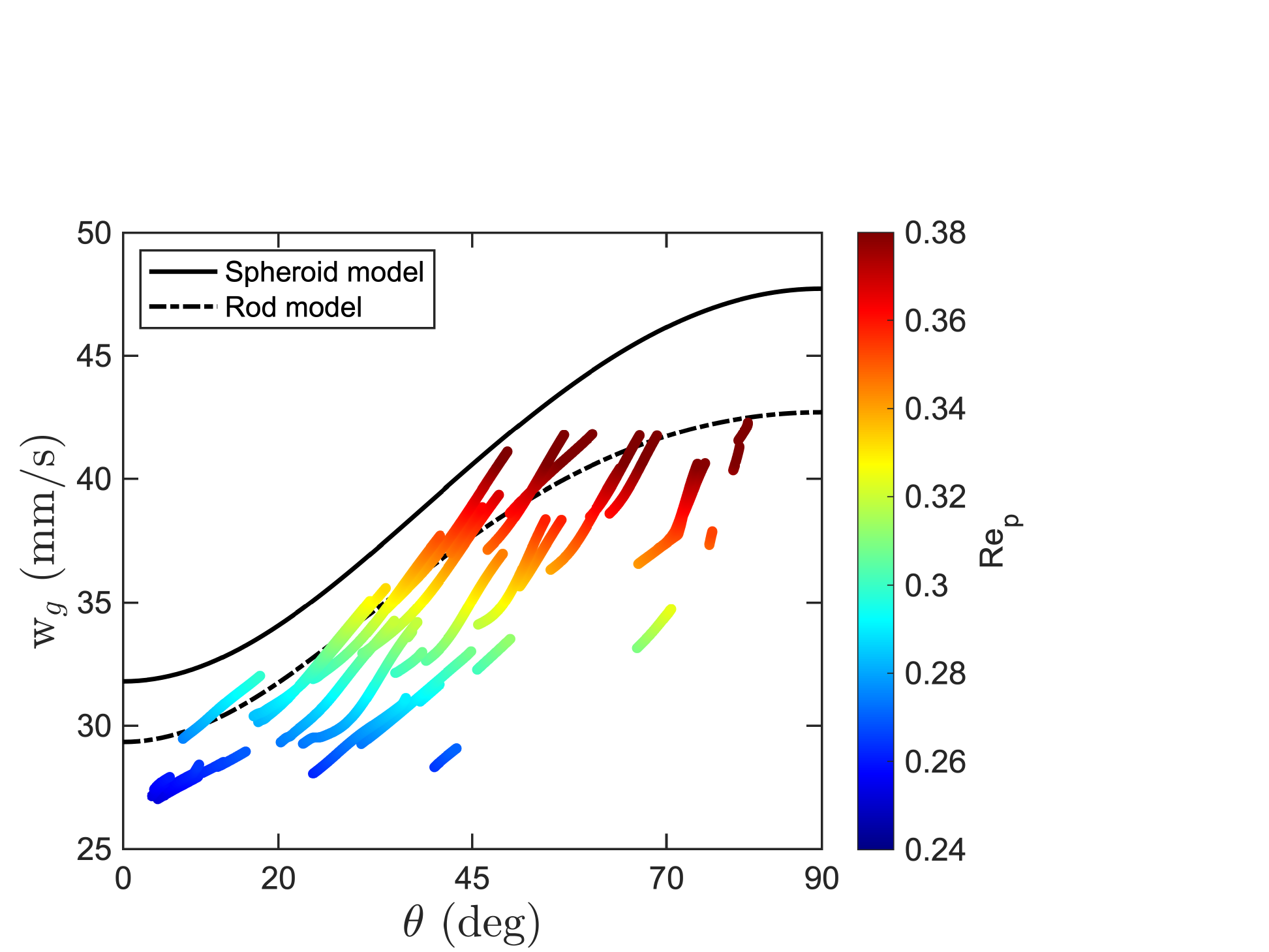}
\put (-130,155) {\large$\beta=16$}
\put (-223,150) {(b)}
%\end{overpic}
\end{subfigure}
 \caption{Evolution of the vertical component of the velocity w$_g$ as a function of the angle $\theta$ for (a) $\beta = 8$ and (b) $\beta = 16$. Each line represents one realization, color-coded by the value of $\mathrm{Re}_p$. Prediction for the spheroid model (solid black line) and for the slender-rod theory~\cite{khayat1989inertia} (dashed-dotted black line).}
  \label{fig:settling_vel}
\end{figure}
%%%%%%%%%%%%%%%%%%%%%%%%%%%%%%%%%%%%%%%%
%%%%%%%%%%%%%%%%%%%%%%%%%%%%%%%%%%%%%%%%
For all experimental realizations, the ratio between the  particle inertial term $m_p \textrm{d}\w/\textrm{d}t$ and the Stokes force term ${\bf f_S}$ in Eq.~(\ref{eq:Newton_x}) was computed to be of order $10^{-3}$.
In other words, the particle inertia contribution in the translational dynamics can be neglected and the particle dynamics is overdamped, as anticipated from the ratio between $\tau_p$ and the characteristic time for the evolution of $\theta$.

Figure~\ref{fig:settling_vel} shows the settling velocity of the rod w$_g$, defined as the projection of ${\bf w}$ on the vertical vector ${\bf g}/|{\bf g}|$, as a function of its orientation $\theta$, for the two values of $\beta$ considered. As for all experimental figures in this article, each trajectory is color-coded with its instantaneous particle Reynolds number $\mathrm{Re}_p$ based on the particle half-length $l$. 
The time series of the evolution of $\theta$ are displayed in Fig.~\ref{fig:angle_vs_time}, and, as anticipated, we observe that the position of the particle, hence its
velocity, evolve over a time scale which is much longer than $\tau_p$. We also observe in Fig.~\ref{fig:settling_vel}
that
the settling velocity is an increasing function of $\theta$: in other words, the settling velocity increases when particles become vertical. This can be readily understood, since the drag decreases when $\theta$ increases from $0$ to $90^\circ$. Therefore, the particle Reynolds number increases linearly with the settling velocity.
The measured settling velocity is compared with the settling velocity calculated by integrating the equation of motion, Eq.~(\ref{eq:Newton_x}), using for the hydrodynamic force the expressions for the spheroids model or for the slender-rod model (see Section~\ref{sec:eq}), and including the particle inertia, despite its weak contribution in the overdamped regime. We recall here that, in the framework of the spheroid model, we use the values of $\beta$ and the mass $m_p$ of the cylindrical particles used in the experiments, and not the values for a spheroid with exactly the same geometrical dimensions as that of the particles used in the experiments, for the computation
of the resistance tensors $\mathbb{A}_S$ and $\mathbb{A}_I$ respectively
due to the Stokes and to the fluid inertia corrections.
The velocity predicted by the spheroid model is shown as a full black line in Fig.~\ref{fig:settling_vel}, and the prediction from the slender-rod model is displayed as a dash-dotted line.

For the two values of $\beta$ considered, the predicted velocities overestimate the measured values, or equivalently, the predicted values of the hydrodynamic forces underestimate the actual force.
However, the qualitative agreement is satisfactory: the relative difference is always smaller than 16\% (for the spheroid model) and 6\% (for the slender-rod model) for $\beta=16$, and respectively 20\% and 10\% for $\beta=8$.
As expected, the agreement with the predictions is better for the slender-rod model than for the spheroid model, as the particles used in the experiment  have a cylindrical shape. Nonetheless, as the aspect ratio increases, the agreement improves from $\beta = 8$ to $\beta = 16$, even with the spheroid model. These observations can be rationalized by noticing that the sharp ends at the extrema of the cylinder play a less important role when $\beta$ increases.

The horizontal component of the measured velocity remains smaller than 20\% of the vertical one, see Figure~\ref{fig:velcomponents} in Appendix~\ref{appendixC}. The figure also shows the ratio of the horizontal and the vertical velocities predicted by the two models discussed in Section~\ref{sec:eq} for spheroids and slender-rods. It can be seen from Figure~\ref{fig:velcomponents} that the ratio of the experimentally measured horizontal and vertical velocities is qualitatively consistent with the predictions of the two models.

\subsection{Angular motion}
\begin{figure}
\centering
\begin{subfigure}
\centering
%\begin{overpic}
\includegraphics[width=0.45\linewidth,trim={0 0 3.5cm 3cm},clip]{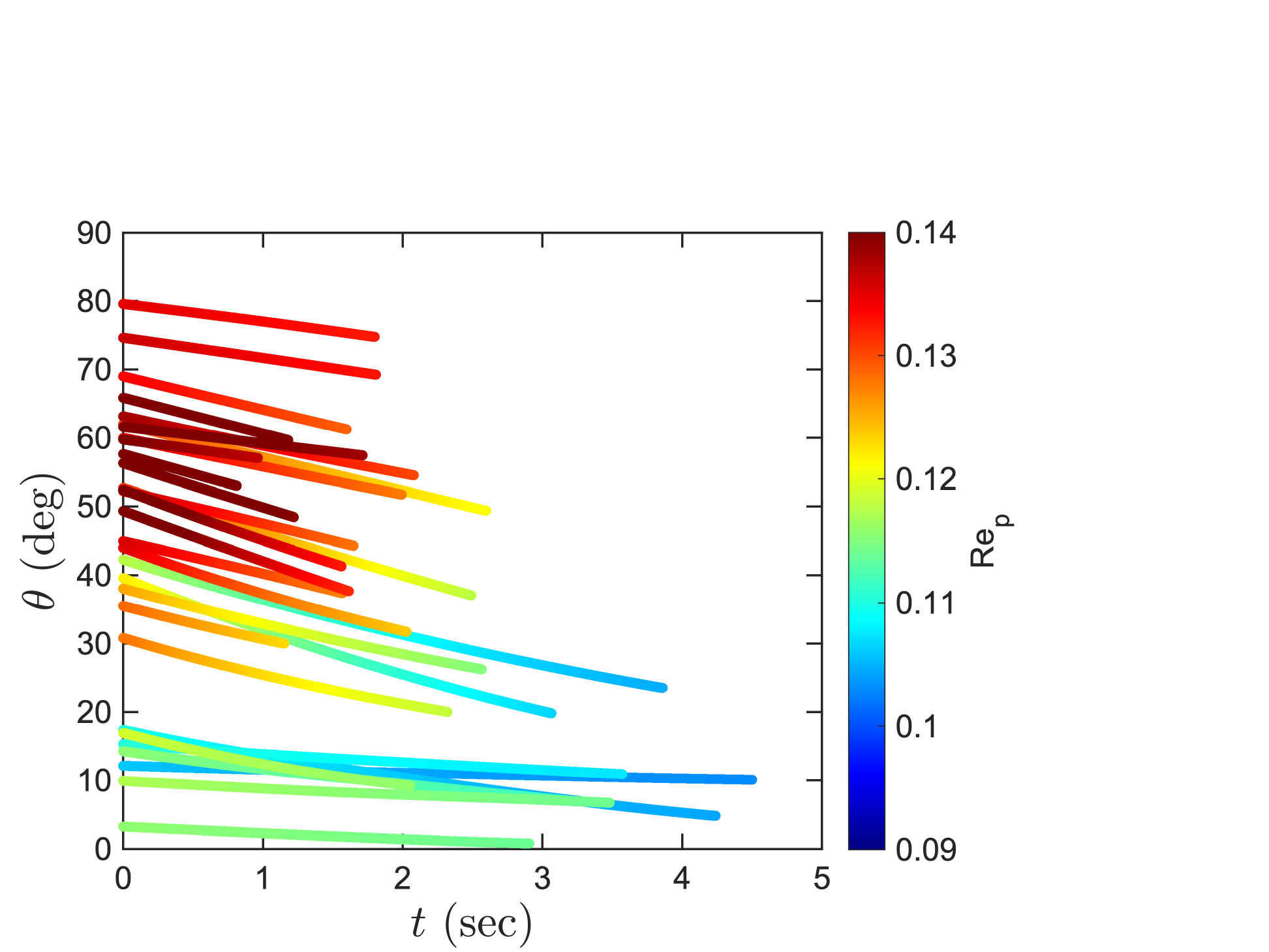}
\put (-130,155) {\large$\beta=8$}
\put (-223,150) {(a)}
%\end{overpic}
\end{subfigure}
\begin{subfigure}
\centering
%\begin{overpic}
\includegraphics[width=0.45\linewidth,trim={0 0 3.5cm 3cm},clip]{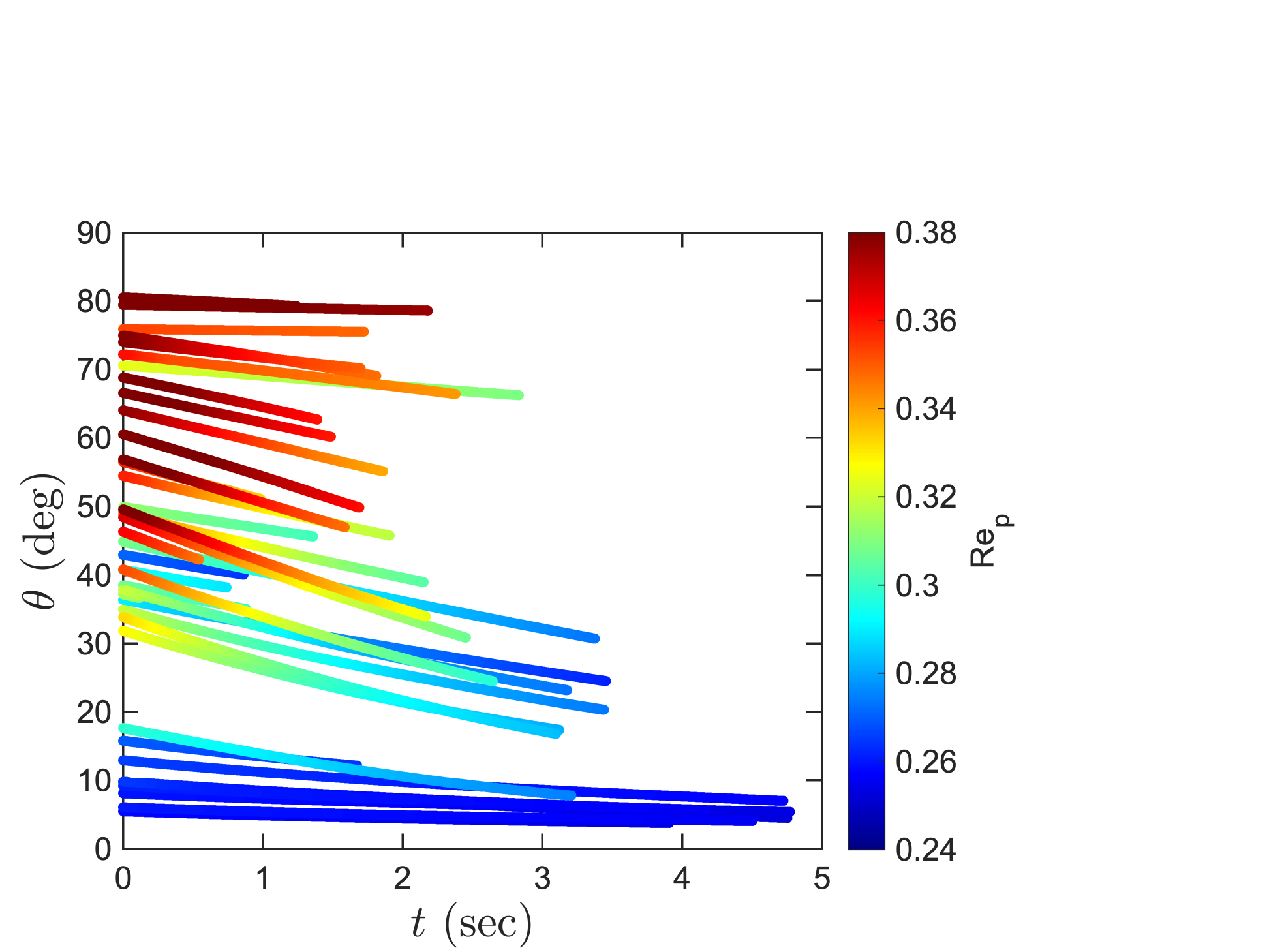}
\put (-130,155) {\large$\beta=16$}
\put (-223,150) {(b)}
%\end{overpic}
\end{subfigure}
\caption{Time evolution of the orientation angle $\theta$ for a) $\beta = 8$ and b) $\beta =16$ for all realizations, color-coded by the value of $\mathrm{Re}_p$. }
\label{fig:angle_vs_time}
\end{figure}

%%%%%%%%%%%%%%%%%%%%%%%%%%%%%%%%%%%%%%%%
%%%%%%%%%%%%%%%%%%%%%%%%%%%%%%%%%%%%%%%%
Figure~\ref{fig:angle_vs_time} shows the time dependence of the angle $\theta$ for the two aspect ratios considered, for a number of different initial rod orientations. The experimental data is color-coded with the instantaneous particle Reynolds number as indicated by the colorbar. The different realizations correspond to different initial orientations covering the range $\theta(t=0)\in (0,\pi/2)$. For all realizations, $\theta$ decreases with time, meaning that particles tend to orient their broadside facing down, resulting in a maximal drag, as previously illustrated in Fig.~\ref{fig:fiber}(c). 
This trend, which has already been reported 
~\cite{jayaweera1965behaviour,bragg1974free,lopez2017inertial,roy2019inertial,gustavsson2019effect,sheikh2020}, is a consequence of the action of the inertial torque $T_I$.  Equation~(\ref{eq:torque_inertial}) shows (for spheroids), in the case of a fluid at rest, that under the effect of this torque, the particle orientation has two fixed points, a vertical one ($\theta=\pi/2$), and an horizontal one ($\theta=0$), but only the latter is stable. This is consistent with the observation that in Fig.~ \ref{fig:angle_vs_time} the magnitude of the angular velocity reduces as the particle approaches horizontal orientation, i.e. as $\theta$ tends to $0$.

Finally, it is worth mentioning that the particle Reynolds number, $\mathrm{Re}_p$, is smaller when the particle settles horizontally ($\theta=0^{\circ}$) as compared to vertically ($\theta=90^{\circ}$). This is a consequence of the fact that the drag exerted on a fiber settling broadside down is higher than that on a fiber settling with its narrow edge first. The later results in a velocity increase with $\theta$, as seen in Figure~\ref{fig:settling_vel}.
%%%%%%%%%%%%%%%%%%%%%%%%%%%%%%%%%%%%%%%%
%%%%%%%%%%%%%%%%%%%%%%%%%%%%%%%%%%%%%%%%
%%%%%%%%%%%%%%%%%%%%%%%%%%%%%%%%%%%%%%%%
%%%%%%%%%%%%%%%%%%%%%%%%%%%%%%%%%%%%%%%%

We now estimate the importance of the particle inertia term in the rotational dynamics, 
Eq.~(\ref{eq:Newton_a}), from the ratio $I_p\ddot{\theta}/C_S\dot{\theta}$. This ratio is very small, well below $10^{-3}$, for the two $\beta$ values.
As a consequence, the left-hand side term of Eq.~(\ref{eq:Newton_a}) is negligible with respect to either of the two terms on the right-hand side. In other words, the particle inertia contribution in the rotational dynamics can be neglected: the fluid-inertia torque ($T_I$) and the Stokes torque ($T_S$) essentially balance each other, so the particle angular dynamics is overdamped. 
Neglecting the second order derivative in Eq.~(\ref{eq:Newton_a}) allows us to simply relate the torque $T_I$ to the instantaneous rotation rate,
$\dot{\theta}$, which therefore provides us with an elementary way to determine $T_I$.
%%%%%%%%%%%%%%%%%%%%%%%%%%%%%%%%%%%%%%%%
%%%%%%%%%%%%%%%%%%%%%%%%%%%%%%%%%%%%%%%%
\begin{figure}[t!]
\centering
\begin{subfigure}
\centering
\includegraphics[width=0.45\linewidth,trim={0 0 3.5cm 3cm},clip]{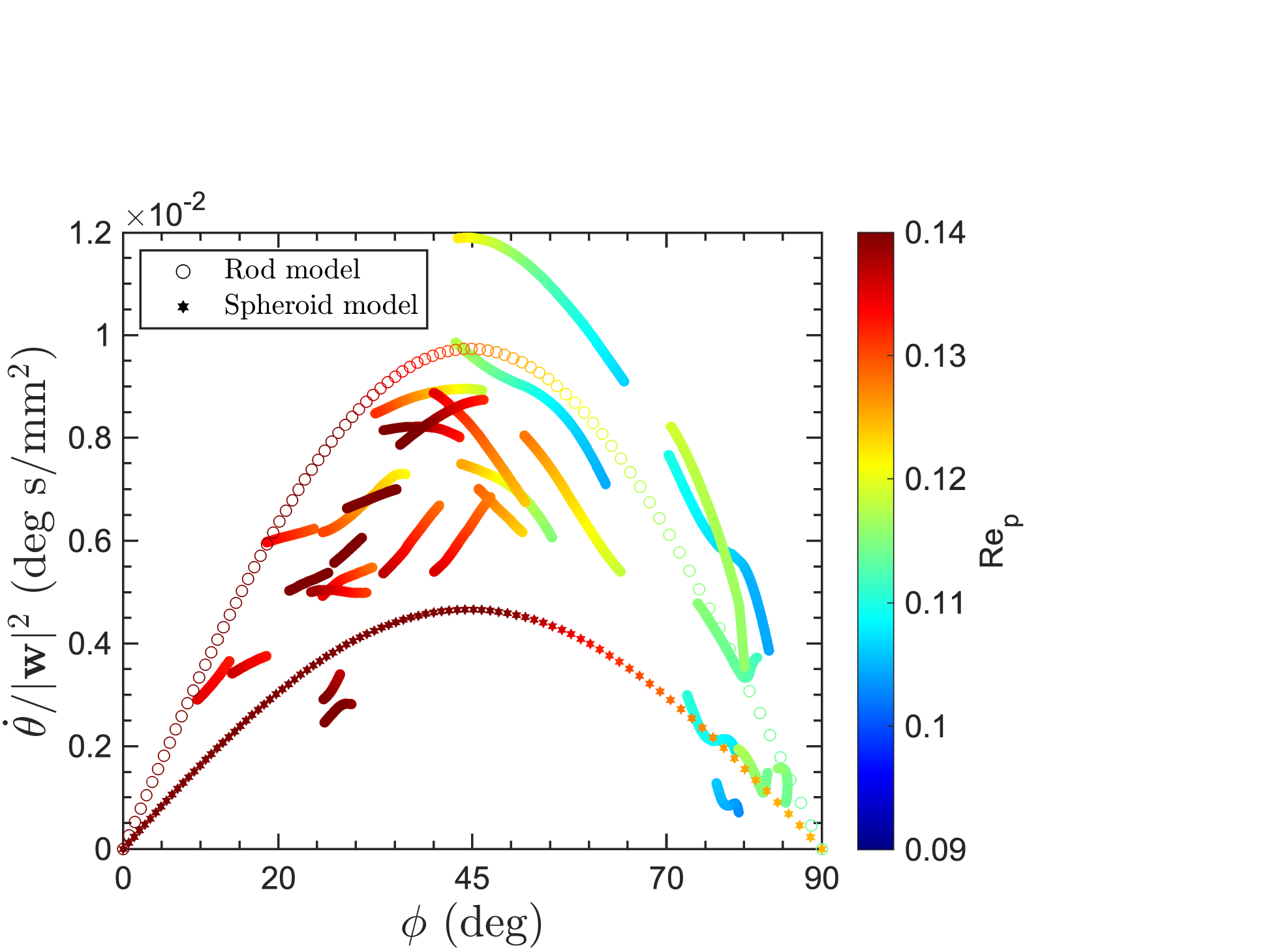}
\put (-130,155) {\large$\beta=8$}
\put (-223,150) {(a)}
\end{subfigure}
\begin{subfigure}
\centering
\includegraphics[width=0.45\linewidth,trim={0 0 3.5cm 3cm},clip]{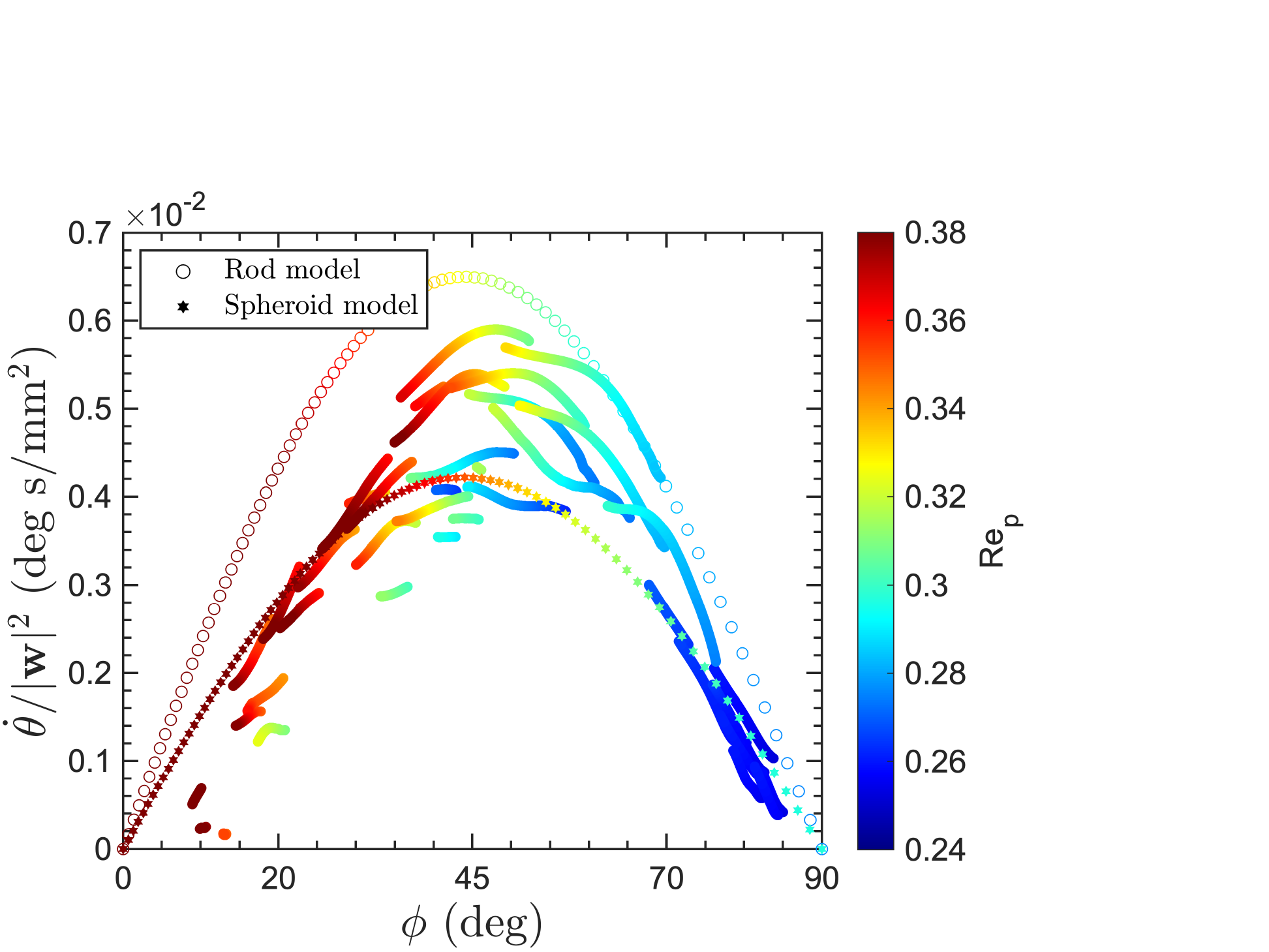}
\put (-130,155) {\large$\beta=16$}
\put (-223,150) {(b)}
\end{subfigure}
\caption{The measured angular velocity, $\dot{\theta}$, divided by $|\w|^2$, color coded by the local value of ${\rm Re}_p$. The predictions of Eq.~\eqref{eq:prediction} are shown for the spheroid model and for the slender-rod model, as indicated by the legend. Panel (a) corresponds to $\beta = 8$, and (b) to $\beta = 16$.}
\label{fig:thetad_vs_phi}
\end{figure}
%%%%%%%%%%%%%%%%%%%%%%%%%%%%%%%%%%%%%%%%
%%%%%%%%%%%%%%%%%%%%%%%%%%%%%%%%%%%%%%%%
To compare the results of the experiments with the prediction of Eq.~\eqref{eq:prediction}, Fig.~\ref{fig:thetad_vs_phi} shows, for $\beta = 8$ (panel a) 
and for $\beta = 16$ (panel b), the measured ratio $\dot{\theta}/|\w|^2$ as
a function of the pitch angle, $\phi$. The trajectories are color coded by the 
recorded value of the particle Reynolds number, ${\rm Re}_p$, as indicated 
by the color bars.  
Equation~\eqref{eq:prediction} predicts that the points should
be along a curve $ \dot{\theta}/| \w|^2 = (C_I/C_S) \sin 2 \phi$. The expected
values for rods (empty circles) and for spheroids (stars) 
are indicated in the figure. We also tested the torque expression proposed by \cite{Kharrouba21}: it leads in the regime of our experiments ($Re_p<0.4$) to curves virtually undistinguishable from those plotted by using the slender-rod model from \cite{khayat1989inertia} (empty circles). As shown in Fig.~\ref{fig:thetad_vs_phi}, the constants $C_I/C_S$ for rods and spheroids differ by a factor $\approx 2$
for $\beta = 8$, and $\approx 1.5$ for $\beta = 16$.

The dependence of $\dot{\theta}/|\w|^2$ qualitatively follows the 
$\sin 2 \phi$ prediction, although with a very large dispersion, particularly
for $\beta = 8$. Interestingly, for both $\beta = 8$ and $\beta = 16$, 
the slender-rod theory overpredicts the value of $\dot{\theta}/|\w|^2$.
On the other hand, the spheroid theory underpredicts the observed dependence,
although the agreement becomes better when $\beta = 16$. 

In view of the limitations of the theory, it is reassuring to see that
the theoretical predictions provide the right order of magnitude for the experimental results.
To discuss further, we recall that the two theories used
here for comparison have different shortcomings. Namely, the slender-rod
theory (open circles) is valid over a larger range of Reynolds numbers (it
only requires that, with our notation, ${\rm Re}_p \ll 1$); but the theory
is valid only when $\log \beta \gg 1$, which is not particularly well 
satisfied. On the other hand, the theory for spheroids works for any 
aspect ratio ($\beta$ does not need to be very large); but the predictions
of the torque deviate by $\sim 30\%$ for values of ${\rm Re}_p = 0.3$, and 
deviations grow when ${\rm Re}_p$ increases~\cite{jiang2020inertial}.
In this context, our observations fall in the delicate regime where both
theories have shortcomings.

The relatively good agreement between the predictions of 
the slender-rod theory~\cite{khayat1989inertia} when $\beta = 8$ is surprising, 
as the value $\beta = 8$ is not particularly large. In fact, the agreement
when $\beta$ increases to $\beta = 16$ tends to deteriorate, which indicates
that the slender-rod approximations does not provide a satisfactory 
description of the results. 
In a similar spirit, the relatively better agreement between our measurements for $\beta = 16$ and the prediction for spheroids is very likely to be coincidental: we do not expect it to  indicate a definite trend, valid at larger values of $\beta$. 

It is interesting to compare the trends in our experiments with previous 
experimental investigations~\cite{roy2019inertial}. As already noticed, the 
expression for the torque obtained in the slender-rod approximation provides a
very good description of our results at $\beta = 8$ (${\rm Re}_p \lesssim 0.15$). 
At the comparable, higher values $\beta = 16$ in our case, 
and $\beta = 20$ in~\cite{roy2019inertial}, the prediction of the
slender-rod theory~\cite{khayat1989inertia}
overpredicts the torque. 
We notice that the Reynolds number in~\cite{roy2019inertial} is significantly 
larger: ${\rm Re}_p \approx 1.6$, although the values of $\rm{Re}_p/\beta$ are
both small. At the larger value of $\beta = 100$ 
in~\cite{roy2019inertial}, the theoretical prediction 
of~\cite{khayat1989inertia} underestimates the experimentally measured torque. 
The Reynolds number, however, is even larger: ${\rm Re}_p \approx 8.6$, 
which adds some extra uncertainties when comparing the measurements with the theoretical expressions, obtained by assuming $\mathrm{Re}_p \ll 1$. 
Note, however, that the value of ${\rm Re}_p/\beta$ remains small: 
${\rm Re}_p/\beta \approx 0.09$. 
For this reason, it would be extremely interesting to obtain results at the large aspect ratios considered in~\cite{roy2019inertial}, but at a lower Reynolds number, as could be obtained by using a more viscous fluid, as done here.

\section{Conclusions}\label{sec:conclusions}
In this work, we studied experimentally the dynamics of rods settling in a quiescent flow. Our experiments were in the regime where the angular dynamics is overdamped, so the torque and the force acting on the object could be readily determined from the settling and from the angular velocities, which was made possible via a 3D-PTV. We considered two particles, with aspect ratios $\beta$ = 8 and $\beta$ = 16. In the two cases, the Reynolds numbers were small, making comparisons with predictions at low particle Reynolds numbers ($\rm Re_p \ll 1$) meaningful.

We compared the measured torques with theoretical predictions in the slender-rod limit, and in the case of spheroids. Our results show that both models qualitatively predict the translational dynamics, in particular the slender-rod theory was found to represent better the experimental data.

Regarding the rod angular dynamics, the models took a simpler form as the rotational dynamics was found to be planar and overdamped. The particles were seen to orient broadside on, i.e. with the maximal drag orientation.

The torque is found to be qualitatively well described by a sin~$2\phi$ functional form, and differences in the torque pre-factor that both models provide were seen. The spheroid theory was found to describe correctly the dynamics of the rods with $\beta$ = 16, whereas for the case $\beta$ = 8 the agreement is less satisfactory, the experimental pre-factor being roughly twice as large as the theory predictions. 
On the other hand, the slender-rod theory happens to quantitatively agree with the data in the case $\beta$ = 8 and, for $\beta$ = 16, tends to over-predict the magnitude of fluid-inertia torque.

\begin{acknowledgements}
The authors are grateful to Vincent Dolique for his help with the Scanning Electron Microscopy. This work was supported by the French research program IDEX-LYON of the University of Lyon in the framework of the French program “Programme Investissements d’Avenir” (ANR-16-IDEX-0005). This research was also supported by Higher Education Commission Pakistan and Campus France.
\end{acknowledgements}
\clearpage
\appendix
\section{Expression for the force acting on anisotropic particles}
\label{appendixA}

We begin by recalling the general expression for the Stokes force:
\begin{equation}
\mathbf{f}_S =  6 \pi \mu a \mathbb{A} (\n) ( \mathbf{u} - \w ),
\end{equation}
where the components of the tensor $\mathbb{A}$ are given by
$A_{ij}  = A_\| n_i n_j + A_\perp ( \delta_{ij} - n_i n_j) $.
The expression for $A_\|$ and $A_\perp$ are, in the case of a spheroid: 
\begin{equation}
   A^s_\| =\frac{8}{3}\frac{\beta}{\chi_0+\beta^2\gamma_0}
  ~~~ {\rm and} ~~~  A^s_\perp =\frac{8}{3}\frac{\beta}{\chi_0+\alpha_0},
\end{equation}
with 
\begin{equation}
    \alpha_o=\frac{\beta^2}{\beta^2-1}-\beta\frac{\mathrm{cosh^{-1}}\beta}{(\beta^2-1)^{3/2}} ~~ {\rm ,} ~~
\gamma_o=\frac{-2}{\beta^2-1}+2\beta\frac{\mathrm{cosh^{-1}}\beta}{(\beta^2-1)^{3/2}}
 ~~ {\rm and} ~~   \chi_0=2\beta \frac{\cosh^{-1}\beta}{(\beta^2-1)^{1/2}}.
\end{equation}
In the case of a rod, the expression of the coefficients $A^r_{\perp} $ 
and $A^r_\|$ are:
\begin{equation}
   A^r_\| = \frac{2}{3}\frac{\beta}{\log \beta}
  ~~~ {\rm and} ~~~  A^r_\perp = \frac{4}{3}\frac{\beta}{\log \beta}.
\end{equation}

The correction to the force due to inertial effects
has been derived for spheroids in~\cite{oberbeck1876ueber}.
The expression for the coefficients
$A^s_{I,\perp}$ and $A^s_{I,\|}$ are:
\begin{equation}
A^s_{I,\|} = [ 3 A_\| - (A_\| \cos^2 \phi + A_\perp \sin^2 \phi) ] A_\| ~~~ 
{\rm and } ~~~ 
A^s_{I,\perp} = [ 3 A_\perp - (A_\| \cos^2 \phi + A_\perp \sin^2 \phi) ] A_\perp, \end{equation}
where $\phi=\cos^{-1}({ \hat{\bf w}\cdot\hat{\bf n}})$.

 In the case of a rod, the fluid-inertia correction to the
force, $\mathbf{f}^r_I$, is given by:
\begin{equation}
    \frac{\mathbf{f}^r_I}{2\pi \mu |{\bf w}|l}=
\bigg(\frac{1}{\log\beta}\bigg)^2\bigg[
\label{eq:stokes_rod_force}
\end{equation}
\begin{equation}\nonumber
  \quad  \frac{2\cos\phi\hat{\bf w}-(2-\cos\phi+\cos^2\phi){\bf n}}{2\mathrm{Re}_p(1-\cos\phi)}\bigg\{E_1[\mathrm{Re}_p(1-\cos\phi)]+\log[\mathrm{Re}_p(1-\cos\phi)]+\gamma-\mathrm{Re}_p(1-\cos\phi)\bigg\}
\end{equation}
\begin{equation}\nonumber
    -\frac{2\cos\phi\hat{\bf w}-(2+\cos\phi+\cos^2\phi){\bf n}}{2\mathrm{Re}_p(1+\cos\phi)}\bigg\{E_1[\mathrm{Re}_p(1+\cos\phi)]+\log[\mathrm{Re}_p(1+\cos\phi)]+\gamma-\mathrm{Re}_p(1+\cos\phi)\bigg\}
\end{equation}
\begin{equation}\nonumber
    -[\cos\phi{\bf n}-2\hat{\bf w}]\bigg\{E_1[\mathrm{Re}_p(1-\cos\phi)]+\log(1-\cos\phi)+E_1[\mathrm{Re}_p(1+\cos\phi)]+\log(1+\cos\phi)
\end{equation}
\begin{equation}
+\frac{1-e^{-\mathrm{Re}_p(1-\cos\phi)}}{\mathrm{Re}_p(1-\cos\phi)}+\frac{1-e^{-\mathrm{Re}_p(1+\cos\phi)}}{\mathrm{Re}_p(1+\cos\phi)}+2(\gamma+\log(\mathrm{Re}_p/4))\bigg\}+3\cos\phi{\bf n}-2\hat{\bf w}\bigg], \nonumber %\label{eq:F_I_khayat}
\end{equation}
where $\hat{\bf w}$ is the unit vector along fiber velocity ${\bf w}$, and $\gamma$ is Euler constant.
\bigbreak
\bigbreak

\section{Expression for the torque acting on anisotropic particles}
\label{appendixB}

The crucial parameter for our work is the ratio $C_I/C_S$ in 
Eq.~\eqref{eq:prediction}, where $C_S$ and $C_I$ are defined by 
Eq.~\eqref{eq:T_S} and \eqref{eq:torque_inertial} respectively. 

\bigskip

\paragraph*{Spheroids}
In the case of spheroids, the shape factor, $F(\beta)$, has been determined 
in~\cite{dabade2015effects} (see in particular Eq.~4.1). The function $F(\beta)$
is represented e.g. in Fig.1 of~\cite{jiang2020inertial}. In 
the limit of very large aspect ratio, $\beta \gg 1$, 
$F(\beta) \approx - 5 \pi/[3 (\log \beta)^2 ]$.   \\

\begin{figure}[h]
\centering
\includegraphics[width=6cm]{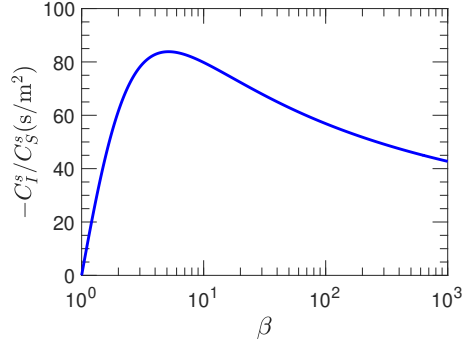}
\caption{The ratio $-C^s_I/C^s_S$ in the case of spheroids, plotted as a function 
of $\beta$ at $\nu=1.05/1216 \mathrm{m^2s^{-1}}$.}
\label{fig:C_I_C_D}
\end{figure}

The ratio between the two coefficients $C_I$ and $C^s_S$, 
appearing in Eq.~\eqref{eq:prediction}, is independent of particle size 
and only depends on $\beta$ and $\nu$; it is given as
\begin{equation}
    \frac{C^s_I}{C^s_S}=\frac{\rho_fl^3F(\beta)}{2}\frac{3}{16}\frac{1}{\pi\mu a^3\beta}\bigg(\frac{\alpha_o+\beta^2\gamma_o}{1+\beta^2}\bigg)=\frac{3}{32}\frac{\beta^2F(\beta)}{\pi\nu}\bigg(\frac{\alpha_o+\beta^2\gamma_o}{1+\beta^2}\bigg)\label{eq:C_I_C_D}
\end{equation}
The ratio in Eq.~\eqref{eq:C_I_C_D} is plotted in figure~\ref{fig:C_I_C_D}.

\paragraph*{Slender-Rods}
In the case of rods, the expression for the torque $T^r_I$ is given by~\cite{khayat1989inertia}:
\begin{equation}
T_I^r = H( {\rm Re}_p, \phi ) |\w| \sin 2 \phi =C^r_I |{\bf w}|^2 \sin 2 \phi,
\end{equation}
where $C^r_I=H(\mathrm{Re_p,\phi})/|{\bf w}|$ with the following expression for $H$:
\begin{equation}\nonumber
\frac{(\log\beta)^2}{2\pi \mu l^2}H(\mathrm{Re_p,\phi})=\frac{1}{2\mathrm{Re}_p(1-\cos\phi)}\bigg\{2+2\frac{e^{-\mathrm{Re}_p(1-\cos\phi)}-1}{\mathrm{Re}_p(1-\cos\phi)}-E_1[\mathrm{Re}_p(1-\cos\phi)]-\log[\mathrm{Re}_p(1-\cos\phi)]-\gamma\bigg\}
\end{equation}
\begin{equation}\nonumber
+\frac{1}{2\mathrm{Re}_p(1+\cos\phi)}\bigg\{2+2\frac{e^{-\mathrm{Re}_p(1+\cos\phi)}-1}{\mathrm{Re}_p(1+\cos\phi)}-E_1[\mathrm{Re}_p(1+\cos\phi)]-\log[\mathrm{Re}_p(1+\cos\phi)]-\gamma\bigg\}
\end{equation}
\begin{equation}
-\frac{1}{\mathrm{Re}_p(1-\cos\phi)\cos\phi}\bigg\{1-\frac{1-e^{-\mathrm{Re}_p(1-\cos\phi)}}{\mathrm{Re}_p(1-\cos\phi)}\bigg\}+\frac{1}{\mathrm{Re}_p(1+\cos\phi)\cos\phi}\bigg\{1-\frac{1-e^{-\mathrm{Re}_p(1+\cos\phi)}}{\mathrm{Re}_p(1+\cos\phi)}\bigg\}.\label{eq:H_factor}
\end{equation}

In the limit $\mathrm{Re}_p \rightarrow 0$, appropriate in the case of our study, and
for $\beta \gg 1$,
we find that:
\begin{equation}
T_I^r \approx - \frac{5 \pi}{6 (\log \beta)^2} \rho_f | \w |^2 l^3 \sin 2 \phi,
\end{equation}
which coincides with the expression derived in~\cite{dabade2015effects}. The ratio $C^r_I/C^r_S$ for rods reads
\begin{equation}
 \frac{C^r_I}{C^r_S}= \frac{H(\mathrm{Re}_p,\phi)}{|{\bf w}|}\frac{3}{8\pi}\frac{\log \beta}{\mu l^3}=\frac{3}{8\pi}\frac{H(\mathrm{Re}_p,\phi)\log \beta}{|{\bf w}|\mu l^3}.
 \label{eq:rod_ratio}
\end{equation}
For a particle settling in a viscous fluid, the ratio in Eq.~\eqref{eq:rod_ratio} is a function of particle orientation through the particle Reynolds number, $\mathrm{Re}_p$, the settling velocity, ${\bf w}$, and the pitch angle, $\phi$. The averaged values (over all orientations) of $C^r_I/C^r_S$ for the cylindrical particles under consideration are -170 and -115 for  $\beta=8$ and $\beta=16$, respectively. The corresponding values in case of spheroids are -82 and -75, see Fig.~\ref{fig:C_I_C_D}.
\section{Horizontal translational dynamics}
\label{appendixC}

\begin{figure}[h!]
\centering
\begin{subfigure}
\centering
%\begin{overpic}
\includegraphics[width=0.45\linewidth,trim={0 0 3.5cm 3cm},clip]{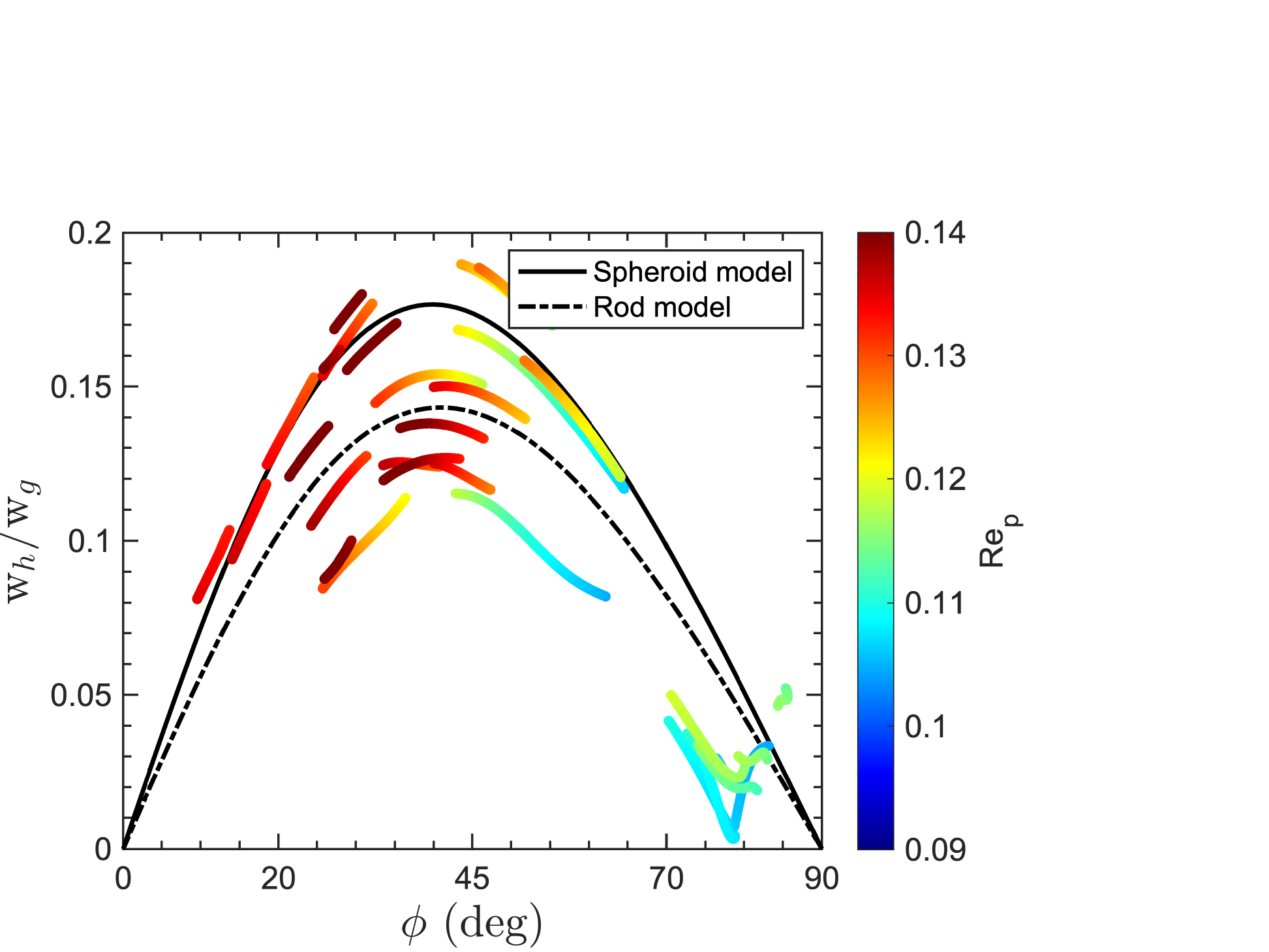}
\put (-130,155) {\large$\beta=8$}
\put (-223,150) {(a)}
%\end{overpic}
\end{subfigure}
\begin{subfigure}
\centering
%\begin{overpic}
\includegraphics[width=0.45\linewidth,trim={0 0 3.5cm 3cm},clip]{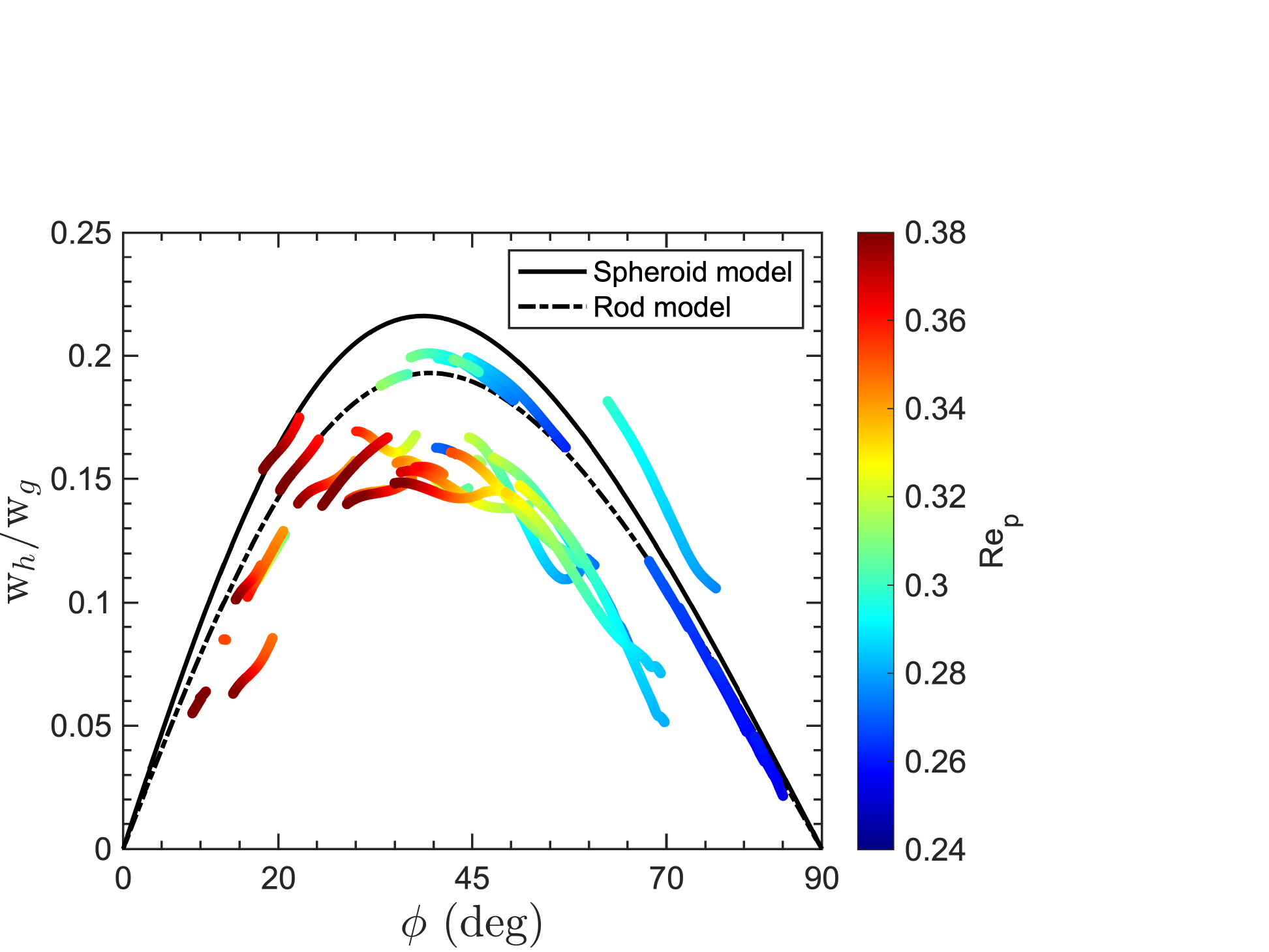}
\put (-130,155) {\large$\beta=16$}
\put (-223,150) {(b)}
%\end{overpic}
\end{subfigure}
\caption{Ratio between horizontal (w$_h$) and vertical (w$_g$) velocity components, as a function of $\phi$.}
\label{fig:velcomponents}
\end{figure}

Figure~\ref{fig:velcomponents} shows the ratio between the horizontal, w$_h$,
and the vertical, w$_g$, components of velocity. As it was the case in 
Fig.~\ref{fig:thetad_vs_phi}, we show the expected ratios from the spheroid
model (full line) and from the slender-rod model (dashed-dotted lines), for
$\beta = 8$ (panel a) and for $\beta = 16$ (panel b).
The ratios are all found to be small, less than $\sim 20 \%$, in agreement
with the predictions from the two models, which differ by no more than 
$\sim 20 \%$ from each other.
%%%%%%%%%%%%%%%%%%%%%%%%%%%%%%%%%%%%%%%%
%%%%%%%%%%%%%%%%%%%%%%%%%%%%%%%%%%%%%%%%
%%%%%%%%%%%%%%%%%%%%%%%%%%%%%%%%%%%%%%%%
%%%%%%%%%%%%%%%%%%%%%%%%%%%%%%%%%%%%%%%%

\bibliography{references}

\begin{thebibliography}{39}%
\makeatletter
\providecommand \@ifxundefined [1]{%
 \@ifx{#1\undefined}
}%
\providecommand \@ifnum [1]{%
 \ifnum #1\expandafter \@firstoftwo
 \else \expandafter \@secondoftwo
 \fi
}%
\providecommand \@ifx [1]{%
 \ifx #1\expandafter \@firstoftwo
 \else \expandafter \@secondoftwo
 \fi
}%
\providecommand \natexlab [1]{#1}%
\providecommand \enquote  [1]{``#1''}%
\providecommand \bibnamefont  [1]{#1}%
\providecommand \bibfnamefont [1]{#1}%
\providecommand \citenamefont [1]{#1}%
\providecommand \href@noop [0]{\@secondoftwo}%
\providecommand \href [0]{\begingroup \@sanitize@url \@href}%
\providecommand \@href[1]{\@@startlink{#1}\@@href}%
\providecommand \@@href[1]{\endgroup#1\@@endlink}%
\providecommand \@sanitize@url [0]{\catcode `\\12\catcode `\$12\catcode
  `\&12\catcode `\#12\catcode `\^12\catcode `\_12\catcode `\%12\relax}%
\providecommand \@@startlink[1]{}%
\providecommand \@@endlink[0]{}%
\providecommand \url  [0]{\begingroup\@sanitize@url \@url }%
\providecommand \@url [1]{\endgroup\@href {#1}{\urlprefix }}%
\providecommand \urlprefix  [0]{URL }%
\providecommand \Eprint [0]{\href }%
\providecommand \doibase [0]{http://dx.doi.org/}%
\providecommand \selectlanguage [0]{\@gobble}%
\providecommand \bibinfo  [0]{\@secondoftwo}%
\providecommand \bibfield  [0]{\@secondoftwo}%
\providecommand \translation [1]{[#1]}%
\providecommand \BibitemOpen [0]{}%
\providecommand \bibitemStop [0]{}%
\providecommand \bibitemNoStop [0]{.\EOS\space}%
\providecommand \EOS [0]{\spacefactor3000\relax}%
\providecommand \BibitemShut  [1]{\csname bibitem#1\endcsname}%
\let\auto@bib@innerbib\@empty
%</preamble>
\bibitem [{\citenamefont {Cox}(1965)}]{cox1965steady}%
  \BibitemOpen
  \bibfield  {author} {\bibinfo {author} {\bibfnamefont {R.~G.}\ \bibnamefont
  {Cox}},\ }\bibfield  {title} {\enquote {\bibinfo {title} {The steady motion
  of a particle of arbitrary shape at small reynolds numbers},}\ }\href@noop {}
  {\bibfield  {journal} {\bibinfo  {journal} {J. Fluid Mech.}\ }\textbf
  {\bibinfo {volume} {23}},\ \bibinfo {pages} {625--643} (\bibinfo {year}
  {1965})}\BibitemShut {NoStop}%
\bibitem [{\citenamefont {Khayat}\ and\ \citenamefont
  {Cox}(1989)}]{khayat1989inertia}%
  \BibitemOpen
  \bibfield  {author} {\bibinfo {author} {\bibfnamefont {R.~E.}\ \bibnamefont
  {Khayat}}\ and\ \bibinfo {author} {\bibfnamefont {R.~G.}\ \bibnamefont
  {Cox}},\ }\bibfield  {title} {\enquote {\bibinfo {title} {Inertia effects on
  the motion of long slender bodies},}\ }\href@noop {} {\bibfield  {journal}
  {\bibinfo  {journal} {J. Fluid Mech.}\ }\textbf {\bibinfo {volume} {209}},\
  \bibinfo {pages} {435--462} (\bibinfo {year} {1989})}\BibitemShut {NoStop}%
\bibitem [{\citenamefont {Dabade}\ \emph {et~al.}(2015)\citenamefont {Dabade},
  \citenamefont {Marath},\ and\ \citenamefont
  {Subramanian}}]{dabade2015effects}%
  \BibitemOpen
  \bibfield  {author} {\bibinfo {author} {\bibfnamefont {V.}~\bibnamefont
  {Dabade}}, \bibinfo {author} {\bibfnamefont {N.~K}\ \bibnamefont {Marath}}, \
  and\ \bibinfo {author} {\bibfnamefont {G.}~\bibnamefont {Subramanian}},\
  }\bibfield  {title} {\enquote {\bibinfo {title} {Effects of inertia and
  viscoelasticity on sedimenting anisotropic particles},}\ }\href@noop {}
  {\bibfield  {journal} {\bibinfo  {journal} {J. Fluid Mech.}\ }\textbf
  {\bibinfo {volume} {778}},\ \bibinfo {pages} {133} (\bibinfo {year}
  {2015})}\BibitemShut {NoStop}%
\bibitem [{\citenamefont {H\"olzer}\ and\ \citenamefont
  {Sommerfeld}(2009)}]{Hoelzer09}%
  \BibitemOpen
  \bibfield  {author} {\bibinfo {author} {\bibfnamefont {A.}~\bibnamefont
  {H\"olzer}}\ and\ \bibinfo {author} {\bibfnamefont {M.}~\bibnamefont
  {Sommerfeld}},\ }\bibfield  {title} {\enquote {\bibinfo {title} {Lattice
  boltzmann simulations to determine drag, lift and torque acting on
  non-spherical particles},}\ }\href@noop {} {\bibfield  {journal} {\bibinfo
  {journal} {Comput. Fluids}\ }\textbf {\bibinfo {volume} {38}},\ \bibinfo
  {pages} {572} (\bibinfo {year} {2009})}\BibitemShut {NoStop}%
\bibitem [{\citenamefont {Zastawny}\ \emph {et~al.}(2012)\citenamefont
  {Zastawny}, \citenamefont {Mallouppas}, \citenamefont {Zhao},\ and\
  \citenamefont {{van Wachem}}}]{Zastawny12}%
  \BibitemOpen
  \bibfield  {author} {\bibinfo {author} {\bibfnamefont {M.}~\bibnamefont
  {Zastawny}}, \bibinfo {author} {\bibfnamefont {G.}~\bibnamefont
  {Mallouppas}}, \bibinfo {author} {\bibfnamefont {F.}~\bibnamefont {Zhao}}, \
  and\ \bibinfo {author} {\bibfnamefont {B.}~\bibnamefont {{van Wachem}}},\
  }\bibfield  {title} {\enquote {\bibinfo {title} {Derivation of drag and lift
  forces and torque coefficients for non-spherical particles in a flow},}\
  }\href@noop {} {\bibfield  {journal} {\bibinfo  {journal} {Int. J. Multiphase
  Flow}\ }\textbf {\bibinfo {volume} {39}},\ \bibinfo {pages} {227} (\bibinfo
  {year} {2012})}\BibitemShut {NoStop}%
\bibitem [{\citenamefont {Jiang}\ \emph {et~al.}(2014)\citenamefont {Jiang},
  \citenamefont {Gallardo},\ and\ \citenamefont {Andersson}}]{Jiang14}%
  \BibitemOpen
  \bibfield  {author} {\bibinfo {author} {\bibfnamefont {F.}~\bibnamefont
  {Jiang}}, \bibinfo {author} {\bibfnamefont {J.~P.}\ \bibnamefont {Gallardo}},
  \ and\ \bibinfo {author} {\bibfnamefont {H.~I.}\ \bibnamefont {Andersson}},\
  }\bibfield  {title} {\enquote {\bibinfo {title} {The laminar wake behind a
  6:1 prolage spheroid at $45^\circ$ incidence angle},}\ }\href@noop {}
  {\bibfield  {journal} {\bibinfo  {journal} {Phys. Fluids}\ }\textbf {\bibinfo
  {volume} {26}},\ \bibinfo {pages} {113601} (\bibinfo {year}
  {2014})}\BibitemShut {NoStop}%
\bibitem [{\citenamefont {Ouchene}\ \emph {et~al.}(2015)\citenamefont
  {Ouchene}, \citenamefont {Khalij}, \citenamefont {Tani\'ere},\ and\
  \citenamefont {Arcen}}]{Ouchene15}%
  \BibitemOpen
  \bibfield  {author} {\bibinfo {author} {\bibfnamefont {R.}~\bibnamefont
  {Ouchene}}, \bibinfo {author} {\bibfnamefont {M.}~\bibnamefont {Khalij}},
  \bibinfo {author} {\bibfnamefont {A.}~\bibnamefont {Tani\'ere}}, \ and\
  \bibinfo {author} {\bibfnamefont {B.}~\bibnamefont {Arcen}},\ }\bibfield
  {title} {\enquote {\bibinfo {title} {Lattice boltzmann simulations to
  determine drag, lift and torque acting on non-spherical particles},}\
  }\href@noop {} {\bibfield  {journal} {\bibinfo  {journal} {Comput. Fluids}\
  }\textbf {\bibinfo {volume} {113}},\ \bibinfo {pages} {53} (\bibinfo {year}
  {2015})}\BibitemShut {NoStop}%
\bibitem [{\citenamefont {Ouchene}\ \emph {et~al.}(2016)\citenamefont
  {Ouchene}, \citenamefont {Khalij}, \citenamefont {Arcen},\ and\ \citenamefont
  {Tani\'ere}}]{Ouchene16}%
  \BibitemOpen
  \bibfield  {author} {\bibinfo {author} {\bibfnamefont {R.}~\bibnamefont
  {Ouchene}}, \bibinfo {author} {\bibfnamefont {M.}~\bibnamefont {Khalij}},
  \bibinfo {author} {\bibfnamefont {B.}~\bibnamefont {Arcen}}, \ and\ \bibinfo
  {author} {\bibfnamefont {A.}~\bibnamefont {Tani\'ere}},\ }\bibfield  {title}
  {\enquote {\bibinfo {title} {A new set of correlations of drag, lift and
  torque coefficients for non-spherical particles at large reynolds numbers},}\
  }\href@noop {} {\bibfield  {journal} {\bibinfo  {journal} {Powder Technol.}\
  }\textbf {\bibinfo {volume} {303}},\ \bibinfo {pages} {33} (\bibinfo {year}
  {2016})}\BibitemShut {NoStop}%
\bibitem [{\citenamefont {Andersson}\ and\ \citenamefont
  {Jiang}(2019)}]{Andersson19}%
  \BibitemOpen
  \bibfield  {author} {\bibinfo {author} {\bibfnamefont {H.~I.}\ \bibnamefont
  {Andersson}}\ and\ \bibinfo {author} {\bibfnamefont {F.}~\bibnamefont
  {Jiang}},\ }\bibfield  {title} {\enquote {\bibinfo {title} {Forces and
  torques on a prolate spheroid: Low-reynolds number and attack angle
  effects},}\ }\href@noop {} {\bibfield  {journal} {\bibinfo  {journal} {Acta
  Mech.}\ }\textbf {\bibinfo {volume} {230}},\ \bibinfo {pages} {431} (\bibinfo
  {year} {2019})}\BibitemShut {NoStop}%
\bibitem [{\citenamefont {Fr\"ohlich}\ \emph {et~al.}(2020)\citenamefont
  {Fr\"ohlich}, \citenamefont {Meinke},\ and\ \citenamefont
  {Schr\"oder}}]{Froehlich20}%
  \BibitemOpen
  \bibfield  {author} {\bibinfo {author} {\bibfnamefont {K.}~\bibnamefont
  {Fr\"ohlich}}, \bibinfo {author} {\bibfnamefont {M.}~\bibnamefont {Meinke}},
  \ and\ \bibinfo {author} {\bibfnamefont {W.}~\bibnamefont {Schr\"oder}},\
  }\bibfield  {title} {\enquote {\bibinfo {title} {Derivation of drag and lift
  forces and torque coefficients for non-spherical particles in a flow},}\
  }\href@noop {} {\bibfield  {journal} {\bibinfo  {journal} {J. Fluid Mech.}\
  }\textbf {\bibinfo {volume} {901}},\ \bibinfo {pages} {A5} (\bibinfo {year}
  {2020})}\BibitemShut {NoStop}%
\bibitem [{\citenamefont {Jiang}\ \emph {et~al.}(2021)\citenamefont {Jiang},
  \citenamefont {Zhao}, \citenamefont {Andersson}, \citenamefont {Gustavsson},
  \citenamefont {Pumir},\ and\ \citenamefont {Mehlig}}]{jiang2020inertial}%
  \BibitemOpen
  \bibfield  {author} {\bibinfo {author} {\bibfnamefont {F}~\bibnamefont
  {Jiang}}, \bibinfo {author} {\bibfnamefont {L}~\bibnamefont {Zhao}}, \bibinfo
  {author} {\bibfnamefont {H}~\bibnamefont {Andersson}}, \bibinfo {author}
  {\bibfnamefont {K}~\bibnamefont {Gustavsson}}, \bibinfo {author}
  {\bibfnamefont {A}~\bibnamefont {Pumir}}, \ and\ \bibinfo {author}
  {\bibfnamefont {B}~\bibnamefont {Mehlig}},\ }\bibfield  {title} {\enquote
  {\bibinfo {title} {Inertial torque on a small spheroid in a uniform flow},}\
  }\href@noop {} {\bibfield  {journal} {\bibinfo  {journal} {Phys. Rev.
  Fluids}\ }\textbf {\bibinfo {volume} {6}} (\bibinfo {year}
  {2021})}\BibitemShut {NoStop}%
\bibitem [{\citenamefont {Kharrouba}\ \emph {et~al.}(2021)\citenamefont
  {Kharrouba}, \citenamefont {Pierson},\ and\ \citenamefont
  {Magnaudet}}]{Kharrouba21}%
  \BibitemOpen
  \bibfield  {author} {\bibinfo {author} {\bibfnamefont {M.}~\bibnamefont
  {Kharrouba}}, \bibinfo {author} {\bibfnamefont {J.-L.}\ \bibnamefont
  {Pierson}}, \ and\ \bibinfo {author} {\bibfnamefont {J.}~\bibnamefont
  {Magnaudet}},\ }\bibfield  {title} {\enquote {\bibinfo {title} {Flow
  structure and loads over inclined cylindrical rodlike particles and
  fibers},}\ }\href@noop {} {\bibfield  {journal} {\bibinfo  {journal} {Phys.
  Rev. Fluids}\ }\textbf {\bibinfo {volume} {6}},\ \bibinfo {pages} {044308}
  (\bibinfo {year} {2021})}\BibitemShut {NoStop}%
\bibitem [{\citenamefont {Lopez}\ and\ \citenamefont
  {Guazzelli}(2017)}]{lopez2017inertial}%
  \BibitemOpen
  \bibfield  {author} {\bibinfo {author} {\bibfnamefont {D.}~\bibnamefont
  {Lopez}}\ and\ \bibinfo {author} {\bibfnamefont {E.}~\bibnamefont
  {Guazzelli}},\ }\bibfield  {title} {\enquote {\bibinfo {title} {Inertial
  effects on fibers settling in a vortical flow},}\ }\href@noop {} {\bibfield
  {journal} {\bibinfo  {journal} {Phys. Rev. Fluids}\ }\textbf {\bibinfo
  {volume} {2}},\ \bibinfo {pages} {024306} (\bibinfo {year}
  {2017})}\BibitemShut {NoStop}%
\bibitem [{\citenamefont {Roy}\ \emph {et~al.}(2019)\citenamefont {Roy},
  \citenamefont {Hamati}, \citenamefont {Tierney}, \citenamefont {Koch},\ and\
  \citenamefont {Voth}}]{roy2019inertial}%
  \BibitemOpen
  \bibfield  {author} {\bibinfo {author} {\bibfnamefont {A.}~\bibnamefont
  {Roy}}, \bibinfo {author} {\bibfnamefont {R.~J.}\ \bibnamefont {Hamati}},
  \bibinfo {author} {\bibfnamefont {L.}~\bibnamefont {Tierney}}, \bibinfo
  {author} {\bibfnamefont {D.~L.}\ \bibnamefont {Koch}}, \ and\ \bibinfo
  {author} {\bibfnamefont {G.~A.}\ \bibnamefont {Voth}},\ }\bibfield  {title}
  {\enquote {\bibinfo {title} {Inertial torques and a symmetry breaking
  orientational transition in the sedimentation of slender fibres},}\
  }\href@noop {} {\bibfield  {journal} {\bibinfo  {journal} {J. Fluid Mech.}\
  }\textbf {\bibinfo {volume} {875}},\ \bibinfo {pages} {576--596} (\bibinfo
  {year} {2019})}\BibitemShut {NoStop}%
\bibitem [{\citenamefont {Candelier}\ and\ \citenamefont
  {Mehlig}(2016)}]{candelier_mehlig_2016}%
  \BibitemOpen
  \bibfield  {author} {\bibinfo {author} {\bibfnamefont {F.}~\bibnamefont
  {Candelier}}\ and\ \bibinfo {author} {\bibfnamefont {B.}~\bibnamefont
  {Mehlig}},\ }\bibfield  {title} {\enquote {\bibinfo {title} {Settling of an
  asymmetric dumbbell in a quiescent fluid},}\ }\href {\doibase
  10.1017/jfm.2016.350} {\bibfield  {journal} {\bibinfo  {journal} {Journal of
  Fluid Mechanics}\ }\textbf {\bibinfo {volume} {802}},\ \bibinfo {pages}
  {174–185} (\bibinfo {year} {2016})}\BibitemShut {NoStop}%
\bibitem [{\citenamefont {Lundell}\ \emph {et~al.}(2011)\citenamefont
  {Lundell}, \citenamefont {S{\"o}derberg},\ and\ \citenamefont
  {Alfredsson}}]{lundell2011fluid}%
  \BibitemOpen
  \bibfield  {author} {\bibinfo {author} {\bibfnamefont {F.}~\bibnamefont
  {Lundell}}, \bibinfo {author} {\bibfnamefont {L.~D.}\ \bibnamefont
  {S{\"o}derberg}}, \ and\ \bibinfo {author} {\bibfnamefont {P.~H.}\
  \bibnamefont {Alfredsson}},\ }\bibfield  {title} {\enquote {\bibinfo {title}
  {Fluid mechanics of papermaking},}\ }\href@noop {} {\bibfield  {journal}
  {\bibinfo  {journal} {Ann. Rev. Fluid Mechanics}\ }\textbf {\bibinfo {volume}
  {43}},\ \bibinfo {pages} {195--217} (\bibinfo {year} {2011})}\BibitemShut
  {NoStop}%
\bibitem [{\citenamefont {Durham}\ \emph {et~al.}(2013)\citenamefont {Durham},
  \citenamefont {Climent}, \citenamefont {Barry}, \citenamefont {De~Lillo},
  \citenamefont {Boffetta}, \citenamefont {Cencini},\ and\ \citenamefont
  {Stocker}}]{durham2013turbulence}%
  \BibitemOpen
  \bibfield  {author} {\bibinfo {author} {\bibfnamefont {W.~M.}\ \bibnamefont
  {Durham}}, \bibinfo {author} {\bibfnamefont {E.}~\bibnamefont {Climent}},
  \bibinfo {author} {\bibfnamefont {M.}~\bibnamefont {Barry}}, \bibinfo
  {author} {\bibfnamefont {F.}~\bibnamefont {De~Lillo}}, \bibinfo {author}
  {\bibfnamefont {G.}~\bibnamefont {Boffetta}}, \bibinfo {author}
  {\bibfnamefont {M.}~\bibnamefont {Cencini}}, \ and\ \bibinfo {author}
  {\bibfnamefont {R.}~\bibnamefont {Stocker}},\ }\bibfield  {title} {\enquote
  {\bibinfo {title} {Turbulence drives microscale patches of motile
  phytoplankton},}\ }\href@noop {} {\bibfield  {journal} {\bibinfo  {journal}
  {Nature Comm.}\ }\textbf {\bibinfo {volume} {4}},\ \bibinfo {pages} {1--7}
  (\bibinfo {year} {2013})}\BibitemShut {NoStop}%
\bibitem [{\citenamefont {Pedley}\ and\ \citenamefont
  {Kessler}(1992)}]{pedley1992hydrodynamic}%
  \BibitemOpen
  \bibfield  {author} {\bibinfo {author} {\bibfnamefont {T.~J.}\ \bibnamefont
  {Pedley}}\ and\ \bibinfo {author} {\bibfnamefont {J.~O.}\ \bibnamefont
  {Kessler}},\ }\bibfield  {title} {\enquote {\bibinfo {title} {Hydrodynamic
  phenomena in suspensions of swimming microorganisms},}\ }\href@noop {}
  {\bibfield  {journal} {\bibinfo  {journal} {Ann. Rev. Fluid Mechanics}\
  }\textbf {\bibinfo {volume} {24}},\ \bibinfo {pages} {313--358} (\bibinfo
  {year} {1992})}\BibitemShut {NoStop}%
\bibitem [{\citenamefont {Ruiz}\ \emph {et~al.}(2004)\citenamefont {Ruiz},
  \citenamefont {Mac{\'\i}as},\ and\ \citenamefont
  {Peters}}]{ruiz2004turbulence}%
  \BibitemOpen
  \bibfield  {author} {\bibinfo {author} {\bibfnamefont {J.}~\bibnamefont
  {Ruiz}}, \bibinfo {author} {\bibfnamefont {D.}~\bibnamefont {Mac{\'\i}as}}, \
  and\ \bibinfo {author} {\bibfnamefont {F.}~\bibnamefont {Peters}},\
  }\bibfield  {title} {\enquote {\bibinfo {title} {Turbulence increases the
  average settling velocity of phytoplankton cells},}\ }\href@noop {}
  {\bibfield  {journal} {\bibinfo  {journal} {Proceedings of the National
  Academy of Sciences}\ }\textbf {\bibinfo {volume} {101}},\ \bibinfo {pages}
  {17720--17724} (\bibinfo {year} {2004})}\BibitemShut {NoStop}%
\bibitem [{\citenamefont {Pruppacher}\ and\ \citenamefont
  {Klett}(1980)}]{pruppacher1980microphysics}%
  \BibitemOpen
  \bibfield  {author} {\bibinfo {author} {\bibfnamefont {H.~R.}\ \bibnamefont
  {Pruppacher}}\ and\ \bibinfo {author} {\bibfnamefont {J.~D.}\ \bibnamefont
  {Klett}},\ }\bibfield  {title} {\enquote {\bibinfo {title} {Microphysics of
  clouds and precipitation},}\ }\href@noop {} {\bibfield  {journal} {\bibinfo
  {journal} {Nature}\ }\textbf {\bibinfo {volume} {284}},\ \bibinfo {pages}
  {88--88} (\bibinfo {year} {1980})}\BibitemShut {NoStop}%
\bibitem [{\citenamefont {Siewert}\ \emph {et~al.}(2014)\citenamefont
  {Siewert}, \citenamefont {Kunnen}, \citenamefont {Meinke},\ and\
  \citenamefont {Schr{\"o}der}}]{siewert2014orientation}%
  \BibitemOpen
  \bibfield  {author} {\bibinfo {author} {\bibfnamefont {C.}~\bibnamefont
  {Siewert}}, \bibinfo {author} {\bibfnamefont {R.~P.~J.}\ \bibnamefont
  {Kunnen}}, \bibinfo {author} {\bibfnamefont {M.}~\bibnamefont {Meinke}}, \
  and\ \bibinfo {author} {\bibfnamefont {W.}~\bibnamefont {Schr{\"o}der}},\
  }\bibfield  {title} {\enquote {\bibinfo {title} {Orientation statistics and
  settling velocity of ellipsoids in decaying turbulence},}\ }\href@noop {}
  {\bibfield  {journal} {\bibinfo  {journal} {Atmospheric research}\ }\textbf
  {\bibinfo {volume} {142}},\ \bibinfo {pages} {45--56} (\bibinfo {year}
  {2014})}\BibitemShut {NoStop}%
\bibitem [{\citenamefont {Gustavsson}\ \emph {et~al.}(2017)\citenamefont
  {Gustavsson}, \citenamefont {Jucha}, \citenamefont {Naso}, \citenamefont
  {L{\'e}v{\^e}que}, \citenamefont {Pumir},\ and\ \citenamefont
  {Mehlig}}]{gustavsson2017statistical}%
  \BibitemOpen
  \bibfield  {author} {\bibinfo {author} {\bibfnamefont {K.}~\bibnamefont
  {Gustavsson}}, \bibinfo {author} {\bibfnamefont {J.}~\bibnamefont {Jucha}},
  \bibinfo {author} {\bibfnamefont {A.}~\bibnamefont {Naso}}, \bibinfo {author}
  {\bibfnamefont {E.}~\bibnamefont {L{\'e}v{\^e}que}}, \bibinfo {author}
  {\bibfnamefont {A.}~\bibnamefont {Pumir}}, \ and\ \bibinfo {author}
  {\bibfnamefont {B.}~\bibnamefont {Mehlig}},\ }\bibfield  {title} {\enquote
  {\bibinfo {title} {Statistical model for the orientation of nonspherical
  particles settling in turbulence},}\ }\href@noop {} {\bibfield  {journal}
  {\bibinfo  {journal} {Phys. Rev. Lett.}\ }\textbf {\bibinfo {volume} {119}},\
  \bibinfo {pages} {254501} (\bibinfo {year} {2017})}\BibitemShut {NoStop}%
\bibitem [{\citenamefont {Jucha}\ \emph {et~al.}(2018)\citenamefont {Jucha},
  \citenamefont {Naso}, \citenamefont {L{\'e}v{\^e}que},\ and\ \citenamefont
  {Pumir}}]{jucha2018settling}%
  \BibitemOpen
  \bibfield  {author} {\bibinfo {author} {\bibfnamefont {J.}~\bibnamefont
  {Jucha}}, \bibinfo {author} {\bibfnamefont {A.}~\bibnamefont {Naso}},
  \bibinfo {author} {\bibfnamefont {E.}~\bibnamefont {L{\'e}v{\^e}que}}, \ and\
  \bibinfo {author} {\bibfnamefont {A.}~\bibnamefont {Pumir}},\ }\bibfield
  {title} {\enquote {\bibinfo {title} {Settling and collision between small ice
  crystals in turbulent flows},}\ }\href@noop {} {\bibfield  {journal}
  {\bibinfo  {journal} {Phys. Rev. Fluids}\ }\textbf {\bibinfo {volume} {3}},\
  \bibinfo {pages} {014604} (\bibinfo {year} {2018})}\BibitemShut {NoStop}%
\bibitem [{\citenamefont {Gustavsson}\ \emph {et~al.}(2019)\citenamefont
  {Gustavsson}, \citenamefont {Sheikh}, \citenamefont {Lopez}, \citenamefont
  {Naso}, \citenamefont {Pumir},\ and\ \citenamefont
  {Mehlig}}]{gustavsson2019effect}%
  \BibitemOpen
  \bibfield  {author} {\bibinfo {author} {\bibfnamefont {K.}~\bibnamefont
  {Gustavsson}}, \bibinfo {author} {\bibfnamefont {M.~Z.}\ \bibnamefont
  {Sheikh}}, \bibinfo {author} {\bibfnamefont {D.}~\bibnamefont {Lopez}},
  \bibinfo {author} {\bibfnamefont {A.}~\bibnamefont {Naso}}, \bibinfo {author}
  {\bibfnamefont {A.}~\bibnamefont {Pumir}}, \ and\ \bibinfo {author}
  {\bibfnamefont {B.}~\bibnamefont {Mehlig}},\ }\bibfield  {title} {\enquote
  {\bibinfo {title} {Effect of fluid inertia on the orientation of a small
  prolate spheroid settling in turbulence},}\ }\href@noop {} {\bibfield
  {journal} {\bibinfo  {journal} {New Journal of Physics}\ }\textbf {\bibinfo
  {volume} {21}},\ \bibinfo {pages} {083008} (\bibinfo {year}
  {2019})}\BibitemShut {NoStop}%
\bibitem [{\citenamefont {Sheikh}\ \emph {et~al.}(2020)\citenamefont {Sheikh},
  \citenamefont {Gustavsson}, \citenamefont {Lopez}, \citenamefont
  {L\'ev\`eque}, \citenamefont {Mehlig}, \citenamefont {Pumir},\ and\
  \citenamefont {Naso}}]{sheikh2020}%
  \BibitemOpen
  \bibfield  {author} {\bibinfo {author} {\bibfnamefont {M.~Z.}\ \bibnamefont
  {Sheikh}}, \bibinfo {author} {\bibfnamefont {K.}~\bibnamefont {Gustavsson}},
  \bibinfo {author} {\bibfnamefont {D.}~\bibnamefont {Lopez}}, \bibinfo
  {author} {\bibfnamefont {E.}~\bibnamefont {L\'ev\`eque}}, \bibinfo {author}
  {\bibfnamefont {B.}~\bibnamefont {Mehlig}}, \bibinfo {author} {\bibfnamefont
  {A.}~\bibnamefont {Pumir}}, \ and\ \bibinfo {author} {\bibfnamefont
  {A.}~\bibnamefont {Naso}},\ }\bibfield  {title} {\enquote {\bibinfo {title}
  {Importance of fluid inertia for the orientation of spheroids settling in
  turbulent flow},}\ }\href {\doibase 10.1017/jfm.2019.1041} {\bibfield
  {journal} {\bibinfo  {journal} {J. Fluid Mech.}\ }\textbf {\bibinfo {volume}
  {886}},\ \bibinfo {pages} {A9} (\bibinfo {year} {2020})}\BibitemShut
  {NoStop}%
\bibitem [{\citenamefont {Kim}\ and\ \citenamefont {Karrila}(1991)}]{Kim:2005}%
  \BibitemOpen
  \bibfield  {author} {\bibinfo {author} {\bibfnamefont {Sangtae}\ \bibnamefont
  {Kim}}\ and\ \bibinfo {author} {\bibfnamefont {Seppo~J.}\ \bibnamefont
  {Karrila}},\ }\href@noop {} {\emph {\bibinfo {title} {Microhydrodynamics:
  principles and selected applications}}}\ (\bibinfo  {publisher}
  {Butterworth-Heinemann},\ \bibinfo {address} {Boston},\ \bibinfo {year}
  {1991})\BibitemShut {NoStop}%
\bibitem [{\citenamefont {Jeffery}(1922)}]{jeffery1922motion}%
  \BibitemOpen
  \bibfield  {author} {\bibinfo {author} {\bibfnamefont {G.~B.}\ \bibnamefont
  {Jeffery}},\ }\bibfield  {title} {\enquote {\bibinfo {title} {The motion of
  ellipsoidal particles immersed in a viscous fluid},}\ }\href@noop {}
  {\bibfield  {journal} {\bibinfo  {journal} {Proc. Royal Soc. London. Ser. A}\
  }\textbf {\bibinfo {volume} {102}},\ \bibinfo {pages} {161--179} (\bibinfo
  {year} {1922})}\BibitemShut {NoStop}%
\bibitem [{\citenamefont {Anand}\ \emph {et~al.}(2020)\citenamefont {Anand},
  \citenamefont {Ray},\ and\ \citenamefont {Subramanian}}]{Ana20}%
  \BibitemOpen
  \bibfield  {author} {\bibinfo {author} {\bibfnamefont {P.}~\bibnamefont
  {Anand}}, \bibinfo {author} {\bibfnamefont {S.~S.}\ \bibnamefont {Ray}}, \
  and\ \bibinfo {author} {\bibfnamefont {G.}~\bibnamefont {Subramanian}},\
  }\bibfield  {title} {\enquote {\bibinfo {title} {Orientation dynamics of
  sedimenting anisotropic particles in turbulence},}\ }\href@noop {} {\bibfield
   {journal} {\bibinfo  {journal} {Phys. Rev. lett.}\ }\textbf {\bibinfo
  {volume} {125}},\ \bibinfo {pages} {034501} (\bibinfo {year}
  {2020})}\BibitemShut {NoStop}%
\bibitem [{\citenamefont {Subramanian}\ and\ \citenamefont
  {Koch}(2005)}]{subramanian2005inertial}%
  \BibitemOpen
  \bibfield  {author} {\bibinfo {author} {\bibfnamefont {G.}~\bibnamefont
  {Subramanian}}\ and\ \bibinfo {author} {\bibfnamefont {D.~L.}\ \bibnamefont
  {Koch}},\ }\bibfield  {title} {\enquote {\bibinfo {title} {Inertial effects
  on fibre motion in simple shear flow},}\ }\href@noop {} {\bibfield  {journal}
  {\bibinfo  {journal} {J. Fluid Mech.}\ }\textbf {\bibinfo {volume} {535}},\
  \bibinfo {pages} {383} (\bibinfo {year} {2005})}\BibitemShut {NoStop}%
\bibitem [{\citenamefont {Kramel}(2017)}]{Kramel:PhD}%
  \BibitemOpen
  \bibfield  {author} {\bibinfo {author} {\bibfnamefont {S.}~\bibnamefont
  {Kramel}},\ }\emph {\bibinfo {title} {Non-Spherical Particle Dynamics in
  Turbulence}},\ \href@noop {} {Ph.D. thesis},\ \bibinfo  {school} {Wesleyan
  College} (\bibinfo {year} {2017})\BibitemShut {NoStop}%
\bibitem [{\citenamefont {Gustavsson}\ \emph {et~al.}(2021)\citenamefont
  {Gustavsson}, \citenamefont {Sheikh}, \citenamefont {Naso}, \citenamefont
  {Pumir},\ and\ \citenamefont {Mehlig}}]{Gus21}%
  \BibitemOpen
  \bibfield  {author} {\bibinfo {author} {\bibfnamefont {K.}~\bibnamefont
  {Gustavsson}}, \bibinfo {author} {\bibfnamefont {M.~Z.}\ \bibnamefont
  {Sheikh}}, \bibinfo {author} {\bibfnamefont {A.}~\bibnamefont {Naso}},
  \bibinfo {author} {\bibfnamefont {A.}~\bibnamefont {Pumir}}, \ and\ \bibinfo
  {author} {\bibfnamefont {B.}~\bibnamefont {Mehlig}},\ }\bibfield  {title}
  {\enquote {\bibinfo {title} {Effect of particle inertia on the alignment of
  small ice crystals in turbulent clouds},}\ }\href@noop {} {\bibfield
  {journal} {\bibinfo  {journal} {J. Atmos. Sci.}\ }\textbf {\bibinfo {volume}
  {in press}} (\bibinfo {year} {2021})}\BibitemShut {NoStop}%
\bibitem [{\citenamefont {Einarsson}\ \emph {et~al.}(2015)\citenamefont
  {Einarsson}, \citenamefont {Candelier}, \citenamefont {Lundell},
  \citenamefont {Angillela},\ and\ \citenamefont {Mehlig}}]{Einarsson2015}%
  \BibitemOpen
  \bibfield  {author} {\bibinfo {author} {\bibfnamefont {J.}~\bibnamefont
  {Einarsson}}, \bibinfo {author} {\bibfnamefont {F.}~\bibnamefont
  {Candelier}}, \bibinfo {author} {\bibfnamefont {F.}~\bibnamefont {Lundell}},
  \bibinfo {author} {\bibfnamefont {J.~R.}\ \bibnamefont {Angillela}}, \ and\
  \bibinfo {author} {\bibfnamefont {B.}~\bibnamefont {Mehlig}},\ }\bibfield
  {title} {\enquote {\bibinfo {title} {Rotation of a spheroid in a simple
  shearat small reynolds number},}\ }\href@noop {} {\bibfield  {journal}
  {\bibinfo  {journal} {Phys. Rev. Lett.}\ }\textbf {\bibinfo {volume} {27}},\
  \bibinfo {pages} {063301} (\bibinfo {year} {2015})}\BibitemShut {NoStop}%
\bibitem [{\citenamefont {Chhabra}\ \emph {et~al.}(2003)\citenamefont
  {Chhabra}, \citenamefont {Agarwal},\ and\ \citenamefont
  {Chaudhary}}]{walleffect}%
  \BibitemOpen
  \bibfield  {author} {\bibinfo {author} {\bibfnamefont {R.~P.}\ \bibnamefont
  {Chhabra}}, \bibinfo {author} {\bibfnamefont {S.}~\bibnamefont {Agarwal}}, \
  and\ \bibinfo {author} {\bibfnamefont {K.}~\bibnamefont {Chaudhary}},\
  }\bibfield  {title} {\enquote {\bibinfo {title} {A note on wall effect on the
  terminal falling velocity of a sphere in quiescent newtonian media in
  cylindrical tubes},}\ }\href {\doibase
  https://doi.org/10.1016/S0032-5910(02)00164-X} {\bibfield  {journal}
  {\bibinfo  {journal} {Powder Technology}\ }\textbf {\bibinfo {volume}
  {129}},\ \bibinfo {pages} {53 -- 58} (\bibinfo {year} {2003})}\BibitemShut
  {NoStop}%
\bibitem [{\citenamefont {Bourgoin}\ and\ \citenamefont
  {Huisman}(2020)}]{micaPTV}%
  \BibitemOpen
  \bibfield  {author} {\bibinfo {author} {\bibfnamefont {M.}~\bibnamefont
  {Bourgoin}}\ and\ \bibinfo {author} {\bibfnamefont {S.~G.}\ \bibnamefont
  {Huisman}},\ }\bibfield  {title} {\enquote {\bibinfo {title} {Using
  ray-traversal for 3d particle matching in the context of particle tracking
  velocimetry in fluid mechanics},}\ }\href {\doibase 10.1063/5.0009357}
  {\bibfield  {journal} {\bibinfo  {journal} {Rev. Sci. Instr.}\ }\textbf
  {\bibinfo {volume} {91}},\ \bibinfo {pages} {085105} (\bibinfo {year}
  {2020})},\ \Eprint {http://arxiv.org/abs/https://doi.org/10.1063/5.0009357}
  {https://doi.org/10.1063/5.0009357} \BibitemShut {NoStop}%
\bibitem [{\citenamefont {Huang}\ \emph {et~al.}(1996)\citenamefont {Huang},
  \citenamefont {Chen},\ and\ \citenamefont {Chia}}]{HUANG96}%
  \BibitemOpen
  \bibfield  {author} {\bibinfo {author} {\bibfnamefont {J.~B.}\ \bibnamefont
  {Huang}}, \bibinfo {author} {\bibfnamefont {Z.}~\bibnamefont {Chen}}, \ and\
  \bibinfo {author} {\bibfnamefont {T.~L.}\ \bibnamefont {Chia}},\ }\bibfield
  {title} {\enquote {\bibinfo {title} {Pose determination of a cylinder using
  reprojection transformation},}\ }\href {\doibase
  https://doi.org/10.1016/0167-8655(96)00061-X} {\bibfield  {journal} {\bibinfo
   {journal} {Pattern Recognition Letters}\ }\textbf {\bibinfo {volume} {17}},\
  \bibinfo {pages} {1089 -- 1099} (\bibinfo {year} {1996})}\BibitemShut
  {NoStop}%
\bibitem [{\citenamefont {Toupoint}\ \emph {et~al.}(2019)\citenamefont
  {Toupoint}, \citenamefont {Ern},\ and\ \citenamefont {Roig}}]{toupoint2019}%
  \BibitemOpen
  \bibfield  {author} {\bibinfo {author} {\bibfnamefont {C.}~\bibnamefont
  {Toupoint}}, \bibinfo {author} {\bibfnamefont {P.}~\bibnamefont {Ern}}, \
  and\ \bibinfo {author} {\bibfnamefont {V.}~\bibnamefont {Roig}},\ }\bibfield
  {title} {\enquote {\bibinfo {title} {Kinematics and wake of freely falling
  cylinders at moderate reynolds numbers},}\ }\href {\doibase
  10.1017/jfm.2019.77} {\bibfield  {journal} {\bibinfo  {journal} {J. Fluid
  Mech.}\ }\textbf {\bibinfo {volume} {866}},\ \bibinfo {pages} {82–111}
  (\bibinfo {year} {2019})}\BibitemShut {NoStop}%
\bibitem [{\citenamefont {Jayaweera}\ and\ \citenamefont
  {Mason}(1965)}]{jayaweera1965behaviour}%
  \BibitemOpen
  \bibfield  {author} {\bibinfo {author} {\bibfnamefont {K.~O.~L.~F.}\
  \bibnamefont {Jayaweera}}\ and\ \bibinfo {author} {\bibfnamefont {B.~J.}\
  \bibnamefont {Mason}},\ }\bibfield  {title} {\enquote {\bibinfo {title} {The
  behaviour of freely falling cylinders and cones in a viscous fluid},}\
  }\href@noop {} {\bibfield  {journal} {\bibinfo  {journal} {J. Fluid Mech.}\
  }\textbf {\bibinfo {volume} {22}},\ \bibinfo {pages} {709--720} (\bibinfo
  {year} {1965})}\BibitemShut {NoStop}%
\bibitem [{\citenamefont {Bragg}\ \emph {et~al.}(1974)\citenamefont {Bragg},
  \citenamefont {van Zuiden},\ and\ \citenamefont {Hermance}}]{bragg1974free}%
  \BibitemOpen
  \bibfield  {author} {\bibinfo {author} {\bibfnamefont {G.M.}\ \bibnamefont
  {Bragg}}, \bibinfo {author} {\bibfnamefont {L.}~\bibnamefont {van Zuiden}}, \
  and\ \bibinfo {author} {\bibfnamefont {C.E.}\ \bibnamefont {Hermance}},\
  }\bibfield  {title} {\enquote {\bibinfo {title} {The free fall of cylinders
  at intermediate reynold's numbers},}\ }\href@noop {} {\bibfield  {journal}
  {\bibinfo  {journal} {Atmospheric Environment (1967)}\ }\textbf {\bibinfo
  {volume} {8}},\ \bibinfo {pages} {755--764} (\bibinfo {year}
  {1974})}\BibitemShut {NoStop}%
\bibitem [{\citenamefont {Oberbeck}(1876)}]{oberbeck1876ueber}%
  \BibitemOpen
  \bibfield  {author} {\bibinfo {author} {\bibfnamefont {A.}~\bibnamefont
  {Oberbeck}},\ }\bibfield  {title} {\enquote {\bibinfo {title} {{\"U}ber
  station{\"a}re {F}l{\"u}ssigkeitsbewegungen mit {B}er{\"u}cksichtigung der
  inneren {R}eibung},}\ }\href@noop {} {\bibfield  {journal} {\bibinfo
  {journal} {J. reine angew. Math.}\ }\textbf {\bibinfo {volume} {81}},\
  \bibinfo {pages} {62--80} (\bibinfo {year} {1876})}\BibitemShut {NoStop}%
\end{thebibliography}%


\end{document}